\documentclass[12pt]{article}
\pdfoutput=1

\usepackage{draft} 
\usepackage{hyperref}
\usepackage{graphicx,color,subfig}
\usepackage{cite}
\usepackage{mciteplus}
\usepackage{skak}
\usepackage{empheq}
\usepackage{anyfontsize}
\DeclareFontFamily{OT1}{pzc}{}
\DeclareFontShape{OT1}{pzc}{m}{it}{<-> s * [1.10] pzcmi7t}{}
\DeclareMathAlphabet{\mathpzc}{OT1}{pzc}{m}{it}

\def\be#1\ee{\begin{align}#1\end{align}}

\begin{document}

\unitlength = .8mm

\begin{titlepage}

\begin{center}

\hfill \\
\hfill \\
\vskip 1cm

\title{Genus Two Modular Bootstrap}

\author{Minjae Cho, Scott Collier, Xi Yin}

\address{
Jefferson Physical Laboratory, Harvard University, \\
Cambridge, MA 02138 USA
}

\email{minjaecho@fas.harvard.edu, scollier@g.harvard.edu, xiyin@fas.harvard.edu}

\end{center}

\abstract{
We study the Virasoro conformal block decomposition of the genus two partition function of a two-dimensional CFT by expanding around a $\mathbb{Z}_3$-invariant Riemann surface that is a three-fold cover of the Riemann sphere branched at four points, and explore constraints from genus two modular invariance and unitarity. In particular, we find ``critical surfaces" that constrain the structure constants of a CFT beyond what is accessible via the crossing equation on the sphere.
}

\vfill

\end{titlepage}

\eject

\begingroup
\hypersetup{linkcolor=black}
\tableofcontents
\endgroup

\section{Introduction} 

The conformal bootstrap program in two dimensions aims to classify and solve two-dimensional conformal field theories (CFTs) based on the associativity of the operator product expansion (OPE) and modular invariance \cite{Ferrara:1973yt, polyakov1974nonhamiltonian, Belavin:1984vu}. A complete set of consistency conditions is given by the crossing equations for sphere 4-point functions and modular covariance of the torus 1-point function for all Virasoro primaries in the CFT \cite{Friedan:1986ua, Moore:1988qv}. In practice, while one may obtain nontrivial constraints on a specific OPE by analyzing a specific sphere 4-point function \cite{Lin:2015wcg, Lin:2016gcl}, or on the entire operator spectrum of the CFT by analyzing the torus partition function \cite{Hellerman:2009bu, Keller:2012mr, Friedan:2013cba, Collier:2016cls}, it has been generally difficult to implement these constraints simultaneously.

In this paper, we analyze modular constraints on the genus two partition function of a general unitary CFT. The modular crossing equation for the Virasoro conformal block decomposition of the genus two partition function encodes both the modular covariance of torus 1-point functions for all primaries and the crossing equation for sphere 4-point functions of pairs of identical primaries. It in principle allows us to constrain the structure constants across the entire spectrum of the CFT.

A technical obstacle in carrying out the genus two modular bootstrap has been the difficulty in computing the genus two conformal blocks. Recently in \cite{Cho:2017oxl} we found a computationally efficient recursive representation of arbitrary Virasoro conformal blocks in the plumbing frame, where the Riemann surface is constructed by gluing two-holed discs with $SL(2,\mathbb{C})$ maps. For a general genus two Riemann surface, however, it is rather cumbersome to map the plumbing parameters explicitly to the period matrix elements on which the modular group $Sp(4,\mathbb{Z})$ acts naturally \cite{Mason:2006dk}.

To circumvent this difficulty, let us recall a well-known reformulation of the modular invariance of the genus one partition function. A torus of complex modulus $\tau$ can be represented as the 2-fold cover of the Riemann sphere, branched over four points at 0, 1, $z$, and $\infty$. $\tau$ and $z$ are related by
\ie\label{enome}
\tau = i {K(1-z)\over K(z)},~~~~ K(z) = {}_2F_1({1\over 2},{1\over 2}, 1|z).
\fe
The torus partition function $Z(\tau, \bar\tau)$ is equal, up to a conformal anomaly factor \cite{Lunin:2000yv}, to the sphere 4-point function of $\mathbb{Z}_2$ twist fields of the 2-fold symmetric product CFT, $\left\langle \sigma_2(0)\sigma_2(z, \bar z)\sigma_2(1)\sigma_2(\infty) \right\rangle$. The modular transformation $\tau \to -1/\tau$ corresponds to the crossing transformation $z\to 1-z$. In this way, the modular invariance of the torus partition function takes a similar form as the crossing equation of the sphere 4-point function, except that the sphere 4-point conformal block is replaced by the torus Virasoro character. 

Usually in the numerical implementation, the crossing equation is rewritten in terms of its $(z,\bar z)$-derivatives evaluated at $z=\bar z ={1\over 2}$. While a priori this requires computing the conformal block (the torus character in this example) at generic $z$, one could equivalently compute instead the conformal block at $z={1\over 2}$ with extra insertions of the stress-energy tensor, or more generally Virasoro descendants of the identity operator at a generic position (on either sheet of the 2-fold cover).

Of course, the above reformulation is unnecessary for analyzing the modular invariance of the genus one partition function, as the torus Virasoro character itself is quite simple. However, it becomes very useful for analyzing genus two modular invariance. Let us begin by considering a 1-complex parameter family of $\mathbb{Z}_3$-invariant genus two Riemann surfaces that are 3-fold covers of the Riemann sphere, branched at 0, 1, $z$, and $\infty$. Following \cite{maloney}, we will refer to them as ``Renyi surfaces"; such surfaces have been studied in the context of entanglement entropy \cite{Calabrese:2009ez,Calabrese:2010he}. For instance, the period matrix of the surface is given by
\ie\label{periodm}
\Omega = \begin{pmatrix} 2 & -1 \\ -1 & 2 \end{pmatrix} {i \; {}_2F_1({2\over 3},{1\over 3},1|1-z) \over \sqrt{3} \; {}_2F_1({2\over 3},{1\over 3},1|z)}.
\fe
The genus two partition function of the CFT in question on the Renyi surface is given, up to a conformal anomaly factor, by the sphere 4-point function of $\mathbb{Z}_3$ twist fields in the 3-fold symmetric product CFT, whose conformal block decomposition takes the form
\ie\label{twist4}
\left\langle \sigma_3(0)\bar\sigma_3(z,\bar z)\sigma_3(1)\bar \sigma_3(\infty) \right\rangle = \sum_{i,j,k\in{\cal I}} C_{ijk}^2 {\cal F}_c(h_i,h_j,h_k|z) {\cal F}_c(\tilde h_i,\tilde h_j,\tilde h_k|\bar z).
\fe
Here ${\cal I}$ is the index set that labels all Virasoro primaries of the CFT, $C_{ijk}$ are the structure constants, and ${\cal F}_c(h_1, h_2, h_3|z)$ is the holomorphic genus two Virasoro conformal block in a particular conformal frame, with central charge $c$ and three internal conformal weights $h_1, h_2, h_3$. We will see that ${\cal F}_c$ can be put in the form
\ie\label{classicalcb}
{\cal F}_c(h_1, h_2, h_3|z) = \exp\left[ c {\cal F}^{cl} (z) \right] {\cal G}_c(h_1, h_2, h_3|z),
\fe
where the factor $\exp\left[ c {\cal F}^{cl} (z) \right]$ captures the large $c$ behavior of the conformal block, essentially due to the conformal anomaly. ${\cal G}_c$ is the genus two conformal block in the plumbing frame of \cite{Cho:2017oxl} (with a different parameterization of the moduli) whose $c\to \infty$ limit is {\it finite}. It admits a recursive representation\footnote{In contrast to the form of the recursion formulae presented in \cite{Cho:2017oxl}, here we include the factor $z^{h_1+h_2+h_3}$ in the definition of the blocks, so that the residue coefficients do not depend on $z$.} 
\ie\label{recursivecb}
{\cal G}_c(h_1, h_2, h_3|z) = {\cal G}_{\infty} (h_1, h_2, h_3|z) + \sum_{i=1}^3 \sum_{r\geq 2, s\geq 1} {{\cal A}^{rs}_i(h_1, h_2, h_3) \over c-c_{rs}(h_i)} {\cal G}_{c_{rs}(h_i)}(h_i \to h_i + rs|z),
\fe
where $c_{rs}(h)$ is a value of the central charge at which a primary of weight $h$ has a null descendant at level $rs$, and ${\cal A}_i^{rs}$ are explicitly known functions of the weights.

The $\mathbb{Z}_3$ cyclic permutations of the three sheets are themselves elements of the $Sp(4,\mathbb{Z})$ modular group. A nontrivial $Sp(4,\mathbb{Z})$ involution that commutes with the $\mathbb{Z}_3$ is the transformation $z\to 1-z$. This gives rise to a genus two modular crossing equation,
\ie\label{modc}
\sum_{i,j,k\in{\cal I}} C_{ijk}^2 \left[ {\cal F}_c(h_i,h_j,h_k|z) {\cal F}_c(\tilde h_i,\tilde h_j,\tilde h_k|\bar z) -{\cal F}_c(h_i,h_j,h_k|1-z) {\cal F}_c(\tilde h_i,\tilde h_j,\tilde h_k| 1- \bar z) \right]=0.
\fe
Together with the non-negativity of $C_{ijk}^2$ for unitary theories, this crossing equation now puts nontrivial constraints on the possible sets of structure constants. For instance, we will find examples of critical surfaces $S$ that bound a (typically compact) domain $D$ in the space of triples of conformal weights $(h_1, h_2, h_3; \tilde h_1, \tilde h_2, \tilde h_3)$, such that the structure constants $C_{ijk}$ with $(h_i, h_j, h_k; \tilde h_i, \tilde h_j, \tilde h_k)$ outside the domain $D$ are bounded by those within the domain $D$. In particular, applying this to noncompact unitarity CFTs, one concludes that there must be triples of primaries in the domain $D$ whose structure constants are nonzero.
We emphasize that the existence of a {\it compact} critical surface for the structure constants is a genuinely nontrivial consequence of genus two modular invariance, which does not follow simply from a combination of bounds on spectral gaps in the OPEs (from analyzing the crossing equation of individual sphere 4-point functions) and modular invariance of the torus partition function (which does not know about the structure constants).

The crossing equation for (\ref{twist4}) does not capture the entirety of genus two modular invariance, since the Renyi surfaces lie on a 1 complex dimensional locus (\ref{periodm}) in the 3 complex dimensional moduli space of genus two Riemann surfaces. Instead of considering general deformations of the geometry, equivalently we can again insert stress-energy tensors on the Renyi surface, or more generally insert Virasoro descendants of the identity operator in the twist field correlator (\ref{twist4}) (on any of the three sheets). This will allow us to access the complete set of genus two modular crossing equations, through the conformal block decomposition of (\ref{twist4}) with extra stress-energy tensor insertions, which is computable explicitly as an expansion in $z$ (or better, in terms of the elliptic nome $q=e^{\pi i \tau}$, where $\tau$ is related to $z$ by (\ref{enome})).

Explicit computation of the genus two Virasoro conformal block of the Renyi surface in the twist-field frame will be given in section \ref{conformalblock}. The genus two modular crossing equation will be analyzed in section \ref{crossing}. In particular, we will find critical surfaces for structure constants simply by taking first order derivatives of the modular crossing equation with respect to the moduli around the crossing invariant point. In section \ref{beyondz}, we formulate the crossing equation beyond the $\mathbb{Z}_3$-invariant locus in the moduli space of genus two Riemann surfaces. We conclude with some future prospectives in section \ref{discussion}.

{\it Note added:} This paper is submitted in coordination with \cite{Cardy:2017qhl} and \cite{Keller:2017iql}, which explore related aspects of two-dimensional conformal bootstrap at genus two.

\section{The genus two conformal block}
\label{conformalblock}

In this section, we will study the genus two Virasoro conformal block with no external operators, focusing on the $\mathbb{Z}_3$-invariant Renyi surface that is a 3-fold branched cover of the Riemann sphere with four branch points. The latter can be represented as the curve
\ie
y^3 = {(x-x_1^+)(x-x_2^+)\over (x-x_1^-)(x-x_2^-)}
\fe
in $\mathbb{P}^1\times \mathbb{P}^1$. The genus two partition function of the CFT on the covering surface can be viewed as a correlation function of the 3-fold symmetric product CFT on the sphere: up to a conformal anomaly factor (dependent on the conformal frame), it is given by the 4-point function of $\mathbb{Z}_3$ twist fields $\sigma_3$ and anti-twist fields $\bar\sigma_3$, $\left\langle \sigma_3(x_1^+) \sigma_3(x_2^+) \bar\sigma_3(x_1^-) \bar\sigma_3(x_2^-) \right\rangle$.
\begin{figure}[h!]
\centering
\subfloat{\includegraphics[width=.4\textwidth]{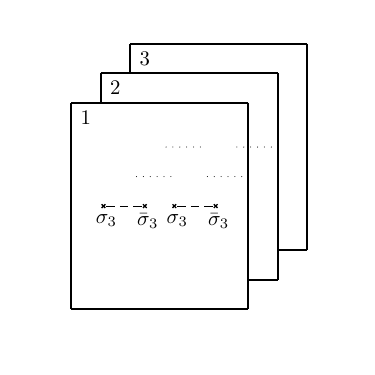}}
\subfloat{\includegraphics[width=.4\textwidth]{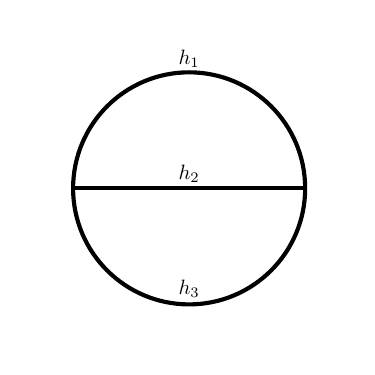}}
\caption{{\bf Left}: The 3-fold cover of the Riemann sphere with four branch points is a genus-two surface. The partition function of the CFT on the covering surface can be regarded as the four-point function of $\bZ_3$ twist fields in the 3-fold product CFT on the sphere. {\bf Right}: The genus two conformal block associated with the $\sigma_3 \bar\sigma_3$ OPE channel.}\label{fig:BranchedCover}
\end{figure}

\subsection{OPE of $\mathbb{Z}_3$-twist fields in ${\rm Sym}^3({\rm CFT})$}

We will begin by analyzing the OPE of the $\mathbb{Z}_3$ twist field $\sigma_3$ and the anti-twist field $\bar\sigma_3$. The 3-fold symmetric product CFT on the sphere with the insertion of $\sigma_3(z_1)$ and $\bar\sigma_3(z_2)$ can be lifted to a single copy of the CFT on the covering space $\Sigma$, which is also a Riemann sphere. Let $t$ be the affine coordinate on the covering sphere. It suffices to consider the special case $z_1=0$, $z_2=1$, where the covering map can be written as
\ie
z = {(t+\omega)^3\over 3\omega (1-\omega) t(t-1)},
\fe
where $\omega = e^{2\pi i /3}$.
The branch points $z_1=0$, $z_2=1$ correspond to $t=-\omega$ and $t=1+\omega$ respectively. We have chosen this covering map (up to $SL(2,\mathbb{C})$ action on $\Sigma$) such that the three points $t_1=0$, $t_2=1$, and $t_3=\infty$ on $\Sigma$ are mapped to $z=\infty$. 

Now let us compute the 3-point function of the pair of twist fields $\sigma_3(0)$, $\bar\sigma_3(1)$, and a general Virasoro descendant operator in the 3-fold tensor product CFT of the form
\ie
\Phi = \bigotimes_{i=1}^3 L_{-N_i} \phi_i
\fe
inserted at $z=\infty$ (as a BPZ conjugate operator). Here we will keep track of the holomorphic $z$-dependence only, and omit the anti-holomorphic sector. For each $i=1,2,3$, $\phi_i$ is a primary of weight $h_i$ in a single copy of the CFT, $N_i=\{n_1^{(i)}, \cdots, n^{(i)}_k\}$ is a partition of the integer $|N_i|$ in descending order, and $L_{-N_i}$ is the Virasoro chain $L_{-n_1^{(i)}}\cdots L_{n_k^{(i)}}$. Following \cite{Lunin:2000yv}, we can write
\ie
\label{thco}
&{\langle \sigma_3(0)\bar\sigma_3(1) \Phi(\infty) \rangle\over \langle \sigma_3(0)\bar\sigma_3(1) \rangle}
= \left\langle {\cal O}_1'(0) {\cal O}_2'(1) {\cal O}_3'(\infty) \right\rangle.
\fe
Here ${\cal O}_i'(t_i)$ is the conformally transformed operator of $L_{-N_i}\phi_i$ on the $i^{\text th}$ covering sheet,\footnote{The factor $z(t_i)^{-2h_i}$ drops out of the correlator (\ref{thco}) due to the normalization convention of $\Phi(\infty)$.}
\ie
{\cal O}_i'(t_i) =(z'(t_i))^{-h_i} {\cal L}^{t_i}_{-N_i} \phi_i'(t_i) = (z(t_i))^{-2h_i} \left[ 3\omega(1-\omega) \right]^{-h_i} {\cal L}^{t_i}_{-N_i} \phi_i'(t_i) ,
\fe
where $\phi_i'(t_i)$ is the corresponding primary in the $t$-frame. ${\cal L}^t_{-N}={\cal L}^t_{-n_1}\cdots {\cal L}^t_{-n_k}$ is the lift of $L_{-N}$ (acting on an operator at $z=\infty$) to the $t$-plane. 
When acting on an operator at $t=t_i$, ${\cal L}^t_{-n}$ is given by
\ie\label{ltn}
{\cal L}^t_{-n} &= -\oint_{C_t} {du\over 2\pi i} {(z(u))^{1+n} \over z'(u)} \left[ T_{uu}(u) - {c\over 12} \{z(u), u\} \right]
\\
&= -\left[3\omega(1-\omega) \right]^{-n}  {\rm Res}_{u\to t} u^{1-n}(u-1)^{1-n} (u+\omega)^{1+3n} (u+\omega^2)^{-2} \left[ T_{uu}(u) - {c\over (u+\omega)^2 (u+\omega^2)^2} \right],
\fe
where we used the Schwarzian derivative 
\ie
\{z,t\} = {12\over (t+\omega)^2 (t+\omega^2)^2}.
\fe
The contour integral in (\ref{ltn}) is taken on the $t$-plane, parameterized by the variable $u$. $C_{t_i}$ is a small counterclockwise circular contour around $t_i$ for $t_1=0$ and $t_2=1$. For $t_3=\infty$, $C_\infty$ is taken to be a large clockwise circular contour on the $t$-plane. Note that the sign convention for the residue at infinity is such that ${\rm Res}_{u\to \infty} {1\over u}=-1$. The overall minus sign on the RHS of (\ref{ltn}) is due to the orientation of the original $z$-contour (where we replace $L_{-n}$ acting on an operator at $z=\infty$ by $L_n$ acting on the product operator $\sigma_3(0)\bar\sigma_3(1)$).

We proceed by putting (\ref{ltn}) into the explicit form
\ie\label{curls}
{\cal L}^t_{-n} = \sum_{m\geq -n} a_{-n,m}^t L_m + c \, b_n^t,
\fe
where
\ie\label{curlt}
& a_{-n,m}^0 = -\left[3\omega(1-\omega) \right]^{-n} {\rm Res}_{u\to 0} u^{-n-m-1}(u-1)^{1-n} (u+\omega)^{1+3n} (u+\omega^2)^{-2} ,
\\
& a_{-n,m}^1 = -\left[3\omega(1-\omega) \right]^{-n} {\rm Res}_{u\to 1} u^{1-n}(u-1)^{-n-m-1} (u+\omega)^{1+3n} (u+\omega^2)^{-2},
\\
& a_{-n,m}^\infty = -\left[3\omega(1-\omega) \right]^{-n} {\rm Res}_{u\to \infty} u^{-n+ m-1}(u-1)^{1-n} (u+\omega)^{1+3n} (u+\omega^2)^{-2} ,
\fe
and
\ie\label{curlb}
& b_n^t = \left[3\omega(1-\omega) \right]^{-n}  {\rm Res}_{u\to t} u^{1-n}(u-1)^{1-n} (u+\omega)^{-1+3n} (u+\omega^2)^{-4} ,
\fe
for $t=0,1,\infty$. On the RHS of (\ref{curls}), $L_m$ is understood to be acting on an operator inserted at $t=0$, 1, or $\infty$.

Putting these together, the 3-point function of interest is
\ie
{}
&{\langle \sigma_3(0)\bar\sigma_3(1) \Phi(\infty) \rangle\over \langle \sigma_3(0)\bar\sigma_3(1) \rangle}
=  C_{123} \left[ 3\omega(1-\omega) \right]^{-h_1-h_2-h_3} \rho( {\cal L}_{-N_3}^\infty h_3, {\cal L}_{-N_2}^1 h_2, {\cal L}_{-N_1}^0 h_1),
\fe
where $C_{123}=\langle \phi_1(0) \phi_2(1) \phi_3(\infty)\rangle$ is the structure constant of the primaries, and $\rho(\xi_3, \xi_2, \xi_1)$ is the 3-point function of Virasoro descendants at $\infty, 1, 0$ on the plane, defined as in \cite{Teschner:2001rv, Cho:2017oxl}. We remind the reader that so far we have only taken into account the holomorphic part of the correlator, for the purpose of deriving the holomorphic Virasoro conformal block in the next section.

\subsection{The conformal block decomposition of $\langle \sigma_3(0) \bar\sigma_3(z) \sigma_3(1)\bar\sigma_3(\infty)\rangle$}\label{cbdecompsec}

Now we turn to the 4-point function of twist fields, $\langle \sigma_3(0) \bar\sigma_3(z) \sigma_3(1)\bar\sigma_3(\infty)\rangle$, and compute the contribution from general untwisted sector descendants of the form $\Phi=\bigotimes_{i=1}^3 L_{-N_i} \phi_i$ in the $\sigma_3(0)\bar\sigma_3(z)$ OPE, for a given triple of primaries $\phi_1, \phi_2, \phi_3$. Again, we focus only on the holomorphic sector. This is given by
\ie\label{lderv}
& \sum_{\{N_i\}, \{M_i\}} \prod_{k=1}^3 G^{N_k M_k}_{h_k} \langle \sigma_3(0) \bar\sigma_3(z) \left[ \otimes_{i=1}^3 L_{-N_i} |\phi_i\rangle \right]  \left[ \otimes_{i=1}^3 \langle \phi_i|L_{-M_i}^\dagger  \right]  \sigma_3(1)\bar\sigma_3(\infty)\rangle
\\
&= \sum_{\{N_i\}, \{M_i\}}  z^{-2h_\sigma + \sum_{i=1}^3(h_i+|N_i|)} \prod_{k=1}^3 G^{N_k M_k}_{h_k} \langle \sigma_3(0) \bar\sigma_3(1) \otimes_{i=1}^3 L_{-N_i} \phi_i(\infty) \rangle \langle \otimes_{j=1}^3 L_{-M_j} \phi_j(0) \sigma_3(1)\bar\sigma_3(\infty)\rangle
\\
&= \sum_{\{N_i\}, \{M_i\}}  z^{-2h_\sigma + \sum_{i=1}^3(h_i+|N_i|)} \prod_{k=1}^3 G^{N_k M_k}_{h_k} \langle \sigma_3(0) \bar\sigma_3(1) \otimes_{i=1}^3 L_{-N_i} \phi_i(\infty) \rangle \langle \bar\sigma_3(0) \sigma_3(1) \otimes_{j=1}^3 L_{-M_j} \phi_j(\infty) \rangle
\\
&= \sum_{\{N_i\}, \{M_i\}}  z^{-2h_\sigma + \sum_{i=1}^3(h_i+|N_i|)} \left[\prod_{k=1}^3 G^{N_k M_k}_{h_k} |z'(t_k)|^{-2h_k} \right] \left\langle  \prod_{i=1}^3 \cL_{-N_i} \phi_i(t_i)  \right\rangle \left\langle  \prod_{j=1}^3 \cL_{-M_j}^* \phi_j(t_j)  \right\rangle .
\fe
Here the summation is over integer partitions in descending order $N_i$ and $M_i$, for $i=1,2,3$, and $G_h^{NM}$ are the inverse Gram matrix elements for a weight $h$ Verma module (nontrivial only for $|N|=|M|$). ${\cal L}_{-M}^*$ is defined as the complex conjugation of ${\cal L}_{-M}$ (not to be confused with the adjoint operator), which simply amounts to replacing $\omega$ by $\omega^2$ in (\ref{curls})-(\ref{curlb}). The appearance of the complex conjugate factors is due to the exchange of $\sigma_3$ with $\bar\sigma_3$ in the last two factors in the third line of (\ref{lderv}). $h_\sigma$ is the holomorphic conformal weight of the $\mathbb{Z}_3$ twist field, given by
\ie
h_\sigma = \left(3-{1\over 3}\right) {c\over 24} = {c\over 9}.
\fe
Using the covering map in the previous section, we arrive at the genus two conformal block for the Renyi surface in the twist field frame
\ie\label{rncb}
{\cal F}_c(h_1,h_2,h_3|z)&= 3^{-3\sum_{i=1}^3 h_i} \sum_{\{N_i\}, \{M_i\}}  z^{-2h_\sigma + \sum_{i=1}^3(h_i+|N_i|)} \prod_{k=1}^3 G^{N_k M_k}_{h_k} 
\\
&~~~\times \rho( {\cal L}_{-N_3}^\infty h_3, {\cal L}_{-N_2}^1 h_2,{\cal L}_{-N_1}^0 h_1) \rho({\cal L}_{-M_3}^{\infty *} h_3, {\cal L}_{-M_2}^{1*} h_2, {\cal L}_{-M_1}^{0*} h_1) ,
\fe
where ${\cal L}_{-N}^{0,1,\infty}$ are given by (\ref{curls})-(\ref{curlb}). 

Let us comment on the $h_i\to 0$ limit, which is rather delicate. If one of the $h_i$ vanishes, say $h_1=0$, corresponding to the vacuum channel in one of the three handles of the genus two surface, then the only conformal blocks that appear in the genus two partition function involve $h_2=h_3$. For $h_2=h_3>0$, the $h_1=0$ block is given by the $h_1\to 0$ limit of (\ref{rncb}). This is not the case however for the vacuum block, where all three weights $h_i$ vanish: in fact the vacuum block differs from the simultaneous $h_i\to 0$ limit of (\ref{rncb}). This is because the latter contains nonvanishing contributions from null descendants of the identity operator that are absent in the vacuum block.

\subsection{Recursive representation}

As already mentioned in the introduction, the genus two conformal block (\ref{rncb}) admits a recursive representation in the central charge of the form (\ref{classicalcb}), (\ref{recursivecb}). The recursion formula is useful in computing the $z$-expansion to high orders efficiently, and can be derived by essentially the same procedure as in \cite{Cho:2017oxl}. The only new feature is that the twist field frame considered here is different from the plumbing frame of \cite{Cho:2017oxl}, which leads to the conformal anomaly factor $\exp\left[ c {\cal F}^{cl}(z) \right]$ in (\ref{classicalcb}). While in principle ${\cal F}^{cl}(z)$ can be determined by evaluating a suitable classical Liouville action as in \cite{Lunin:2000yv}, we find it more convenient to compute ${\cal F}^{cl}(z)$ by directly inspecting the large $c$ limit of $\log {\cal F}_c(h_1, h_2, h_3|z)$. Indeed, the latter is linear in $c$ in the large $c$ limit (with a leading coefficient that is independent of the internal weights), with the following series expansion in $z$
\ie
{\cal F}^{cl}(z) & =-{2\over9}\text{log}(z) +6 \left({z\over 27}\right)^2 + 162 \left({z\over 27}\right)^3 + 3975 \left({z\over 27}\right)^4 + 96552 \left({z\over 27}\right)^5 + 2356039 \left({z\over 27}\right)^6
\\
&~~~+ 57919860 \left({z\over 27}\right)^7  + {2869046823\over 2} \left({z\over 27}\right)^8+35771031918\left({z\over 27}\right)^9+{4486697950566\over 5}\left({z\over 27}\right)^{10}\\
&~~~ +{\cal O}(z^{11}).
\fe
Note that ${c\over 3} {\cal F}^{cl}(z)$ agrees with the semiclassical Virasoro sphere 4-point conformal block of central charge $c$ in the vacuum channel with four external primaries of weight ${h_\sigma\over 3}={c\over 27}$ \cite{Zamolodchikov:426555, Chang:2016ftb}.
For numerical computations, we can pass to the elliptic nome parameter $q=e^{\pi i \tau}$, where $\tau$ is related to $z$ via (\ref{enome}). The $q$-expansion converges much faster than the $z$-expansion\footnote{As explained in \cite{Chang:2016ftb}, the $q$-expansion of ${\cal F}^{cl}$ in general need not have unit radius of convergence, due to possible zeroes of the conformal block. In the present example, the radius of convergence nonetheless appears to be 1. We thank Y.-H. Lin for pointing out this subtlety and providing numerical verifications.} evaluated at the crossing symmetric point $z={1\over 2}$, which corresponds to $q=e^{-\pi}$.

After factoring out $\exp\left[ c {\cal F}^{cl}(z) \right]$, the remaining part of the conformal block ${\cal G}_c(h_1, h_2, h_3|z)$ as a function of the central charge $c$ has poles at 
\ie
c_{rs}(h)=1+6(b_{rs}(h) + b_{rs}(h)^{-1})^2, ~~~ {\rm with}~b_{rs}(h)^2= {rs - 1 + 2h+ \sqrt{(r-s)^2 + 4(rs-1) h + 4h^2}\over 1-r^2},
\fe
where $r\geq 2$ and $s\geq 1$, for $h=h_i$, $i=1,2,3$. The residue at a pole $c=c_{rs}(h_i)$ is proportional to the conformal block with central charge $c_{rs}(h_i)$ and the weight $h_i$ shifted to $h_i+rs$. The precise recursion formula is
\ie\label{gcrec}
{\cal G}_c(h_1, h_2, h_3|z) &= {\cal G}_\infty(h_1, h_2, h_3|z) 
\\
& ~~~ +
\sum_{r\geq 2, s\geq 1}  \left[ - {\partial c_{rs}(h_1)\over \partial h_1} \right] {A_{rs}^{c_{rs}(h_1)} \left(P^{rs}_{c_{rs}(h_1)}\begin{bmatrix} h_2 \\ h_3 \end{bmatrix} \right)^2 \over c-c_{rs}(h_1)}{\cal G}_{c_{rs}(h_i)}(h_i \to h_i + rs|z)
\\
& ~~~ + (2~{\rm cyclic~permutations~on}~ h_1, h_2, h_3),
\fe
where $A_{rs}^c$ is the constant
\ie
A_{rs}^c = {1\over 2} \prod_{m = 1-r}^r \prod_{n=1-s}^s (mb + n b^{-1})^{-1},~~~ (m,n)\not=(0,0),\, (r,s),
\fe
for $c=1+6(b+b^{-1})^2$, and $P^{rs}_c$ is the fusion polynomial
\ie
P_c^{rs} \begin{bmatrix}d_1\\d_2\end{bmatrix} = \prod_{p=1-r~{\rm step}~2}^{r-1} \prod_{q=1-s~{\rm step}~2}^{s-1}
{\lambda_1+\lambda_2 + pb + q b^{-1}\over 2} {\lambda_1-\lambda_2 + pb + q b^{-1}\over 2},
\fe
where $\lambda_i$ are related to the weights $d_i$ by $d_i = {1\over 4}(b+b^{-1})^2 - {1\over 4}\lambda_i^2$.

The remaining undetermined piece in the formula (\ref{gcrec}) is the $c\to \infty$ limit ${\cal G}_\infty(h_1, h_2, h_3|z)$. 
It was shown in \cite{Cho:2017oxl} that ${\cal G}_\infty(h_1, h_2, h_3|z)$ is the product of the vacuum block and $SL(2,\mathbb{C})$ global block in the plumbing frame. The vacuum block is given by the holomorphic part of the gravitational 1-loop free energy of the genus two hyperbolic handlebody, computed in \cite{Giombi:2008vd}. To translate the result of \cite{Giombi:2008vd} into the vacuum part of our ${\cal G}_\infty$ requires expressing the Schottky parameters of the Renyi surface in terms of $z$; this can be achieved through the map between Schottky parameters and the period matrix (\ref{periodm}). Furthermore, the global block of \cite{Cho:2017oxl} is naturally expressed in terms of the plumbing parameters, whose map to $z$ is nontrivial. The implementation of an efficient recursive computational algorithm for the genus two conformal blocks in the twist field frame will require knowing ${\cal G}_\infty$, which is in principle computable given the above ingredients, based on the map from $z$ to the Schottky parameters and the plumbing parameters of the Renyi surface. Here we simply evaluate the $z$-expansion (\ref{rncb}) directly, strip off the conformal anomaly factor and then take the $c\to \infty$ limit, giving the result 
\ie\nonumber
{\cal G}_\infty&(h_1, h_2, h_3|z) = \left({z\over 27}\right)^{h_1+h_2+h_3}\bigg\{ 1 + \left[{h_1+h_2+h_3\over 2}+{(h_2-h_3)^2\over 54h_1} + {(h_3-h_1)^2\over 54 h_2}+{(h_1-h_2)^2\over 54 h_3}\right] z
\fe
{\fontsize{3}{3}
$$\renewcommand*{\arraystretch}{.4} \begin{array}{l}
+{z^2\over 2916 h_1(1+2h_1)h_2 (1+2h_2)h_3(1+2h_3)} \bigg[ 4 h_2^2 h_1^6+4 h_3^2 h_1^6+6 h_2 h_1^6+8 h_2 h_3 h_1^6+6 h_3 h_1^6+2 h_1^6-16 h_2^3 h_1^5-16 h_3^3 h_1^5+94 h_2^2 h_1^5+200 h_2 h_3^2 h_1^5+94 h_3^2 h_1^5 +45 h_2 h_1^5+200 h_2^2 h_3 h_1^5+188 h_2 h_3 h_1^5+45 h_3 h_1^5-3 h_1^5+24 h_2^4 h_1^4+24 h_3^4 h_1^4
-100 h_2^3 h_1^4-208 h_2 h_3^3 h_1^4  -100 h_3^3 h_1^4+118 h_2^2 h_1^4 \\ 
 +3380 h_2^2 h_3^2 h_1^4 +1938 h_2 h_3^2 h_1^4+118 h_3^2 h_1^4+87 h_2 h_1^4-208 h_2^3 h_3 h_1^4 +1938 h_2^2 h_3 h_1^4+1197 h_2 h_3 h_1^4+87 h_3 h_1^4-16 h_2^5 h_1^3-16 h_3^5 h_1^3-100 h_2^4 h_1^3-208 h_2 h_3^4 h_1^3 -100 h_3^4 h_1^3-330 h_2^3 h_1^3+5376 h_2^2 h_3^3 h_1^3+2008 h_2 h_3^3 h_1^3-330 h_3^3 h_1^3-84 h_2^2 h_1^3+5376 h_2^3 h_3^2 h_1^3+11776 h_2^2 h_3^2 h_1^3+4374 h_2 h_3^2 h_1^3
\\
-84 h_3^2 h_1^3+31 h_2 h_1^3 -208 h_2^4 h_3 h_1^3+2008 h_2^3 h_3 h_1^3+4374 h_2^2 h_3 h_1^3+1722 h_2 h_3 h_1^3+31 h_3 h_1^3+h_1^3+4 h_2^6 h_1^2+4 h_3^6 h_1^2+94 h_2^5 h_1^2+200 h_2 h_3^5 h_1^2+94 h_3^5 h_1^2+118 h_2^4 h_1^2 +3380 h_2^2 h_3^4 h_1^2+1938 h_2 h_3^4 h_1^2+118 h_3^4 h_1^2-84 h_2^3 h_1^2+5376 h_2^3 h_3^3 h_1^2+11776 h_2^2 h_3^3 h_1^2+4374 h_2 h_3^3 h_1^2-84 h_3^3 h_1^2-62 h_2^2 h_1^2\\
 +3380 h_2^4 h_3^2 h_1^2 +11776 h_2^3 h_3^2 h_1^2+11148 h_2^2 h_3^2 h_1^2+2926 h_2 h_3^2 h_1^2-62 h_3^2 h_1^2-h_2 h_1^2+200 h_2^5 h_3 h_1^2+1938 h_2^4 h_3 h_1^2+4374 h_2^3 h_3 h_1^2+2926 h_2^2 h_3 h_1^2+597 h_2 h_3 h_1^2 -h_3 h_1^2+6 h_2^6 h_1+8 h_2 h_3^6 h_1+6 h_3^6 h_1+45 h_2^5 h_1+200 h_2^2 h_3^5 h_1+188 h_2 h_3^5 h_1+45 h_3^5 h_1+87 h_2^4 h_1-208 h_2^3 h_3^4 h_1+1938 h_2^2 h_3^4 h_1\\
 +1197 h_2 h_3^4 h_1 +87 h_3^4 h_1+31 h_2^3 h_1-208 h_2^4 h_3^3 h_1+2008 h_2^3 h_3^3 h_1+4374 h_2^2 h_3^3 h_1+1722 h_2 h_3^3 h_1+31 h_3^3 h_1-h_2^2 h_1+200 h_2^5 h_3^2 h_1+1938 h_2^4 h_3^2 h_1+4374 h_2^3 h_3^2 h_1 +2926 h_2^2 h_3^2 h_1+597 h_2 h_3^2 h_1-h_3^2 h_1+8 h_2^6 h_3 h_1+188 h_2^5 h_3 h_1+1197 h_2^4 h_3 h_1+1722 h_2^3 h_3 h_1+597 h_2^2 h_3 h_1+6 h_2 h_3 h_1+2 h_2^6\\
 +4 h_2^2 h_3^6+6 h_2 h_3^6+2 h_3^6-3 h_2^5-16 h_2^3 h_3^5+94 h_2^2 h_3^5+45 h_2 h_3^5-3 h_3^5+24 h_2^4 h_3^4-100 h_2^3 h_3^4+118 h_2^2 h_3^4+87 h_2 h_3^4+h_2^3-16 h_2^5 h_3^3-100 h_2^4 h_3^3-330 h_2^3 h_3^3 -84 h_2^2 h_3^3+31 h_2 h_3^3+h_3^3+4 h_2^6 h_3^2+94 h_2^5 h_3^2+118 h_2^4 h_3^2-84 h_2^3 h_3^2-62 h_2^2 h_3^2-h_2 h_3^2+6 h_2^6 h_3+45 h_2^5 h_3+87 h_2^4 h_3+31 h_2^3 h_3-h_2^2 h_3\bigg]
\end{array}
$$
}
\ie\label{regpart}
+{\cal O}(z^3)\bigg\}.~~~~~~~~~~~~~~~~~~~~~~~~~~~~~~~~~~~~~~~~~~~~~~~~~~~~~~~~~~~~~~~~~~~~~~~~~~~~~~~~~~~~~~~~~~~~
\fe
As already noted, the analog of ${\cal G}_\infty$ for the vacuum block, ${\cal G}_\infty^0(z)$, is not the same as the simultaneous $h_i\to 0$ limit of (\ref{regpart}). The first few terms in the $z$-expansion of ${\cal G}_\infty^0(z)$ is given explicitly by
\ie
{\cal G}_\infty^{0}(z) =& 1 + 3 \left({z\over 27}\right)^4+168\left({z\over 27}\right)^5+6567\left({z\over 27}\right)^6+222012\left({z\over 27}\right)^7+6960036\left({z\over 27}\right)^8+\mathcal{O}(z^9).
\fe

\subsection{Mapping to the 3-fold-pillow}\label{pillowsection}

In this section we consider the Renyi surface in the 3-fold-pillow frame, which makes obvious certain positivity properties of the genus two conformal block. Following \cite{Maldacena:2015iua}, the map from the plane (parameterized by $w$) to the pillow (parameterized by $v$) is given by
\ie\label{pillowmap}
v = {1\over (\theta_3(\tau))^2} \int_0^w {dx\over \sqrt{x(1-x)(z-x)}}.
\fe 
The four branch points on the plane at $0, z, 1, \infty$, where the $\mathbb{Z}_3$ twist fields and anti-twist fields are inserted, are mapped to $v=0, \pi, \pi(\tau+1), \pi\tau$ respectively, where $\tau$ is given by (\ref{enome}). The covering surface is turned into a 3-fold cover of the pillow, with the twist fields inserted at the four corners.
\begin{figure}[h!]
\centering
\subfloat{\includegraphics[width=.45\textwidth]{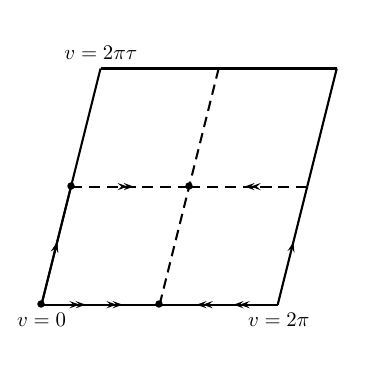}}
\subfloat{\includegraphics[width=.45\textwidth]{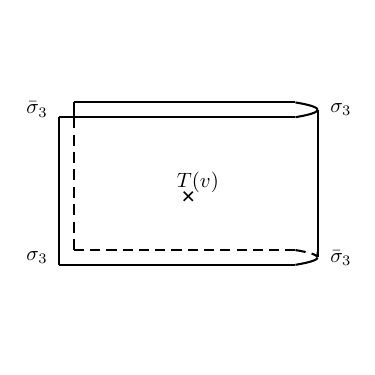}}
\caption{{\bf Left}: The pillow geometry is the quotient $T^2/\bZ_2$. The four branch points on the plane $0,z,1,\infty$ are mapped to the $\bZ_2$ fixed points $v=0,\pi,\pi(\tau+1),\pi\tau$ respectively. {\bf Right}: The pillow with the $\bZ_3$ twist fields inserted at the corners. In Section \ref{beyondz} we will obtain the full set of genus two modular crossing equations by inserting the stress-energy tensor or more generally arbitrary Virasoro descendants of the identity at the front center on each sheet of the 3-fold-pillow.}\label{fig:pillow}
\end{figure}

The Renyi surface conformal block in the twist field frame can be mapped to the pillow frame as
\ie\label{pillowexpansion}
{\cal F}_c(h_1, h_2, h_3|z) &= (z(1-z))^{{c\over 8} - 2 h_\sigma} \theta_3(\tau)^{{3c\over 2} - 16 h_\sigma} q^{{h_1+h_2+h_3}-{c\over 8}} \sum_{n=0}^\infty A_n(h_1, h_2, h_3)  q^{n},
\fe
where $q=e^{\pi i \tau}$, $h_\sigma = {c\over 9}$. For instance, the first few coefficients $A_0$, $A_1$ and $A_2$ are given by
\ie\label{acoeffs}
A_0&= 2^{-{c\over 2}} \left( {16\over 27} \right)^{h_1+h_2+h_3},~~~~~~
\\
A_1&= 2^{-{c\over 2}-1} \left( {16\over 27} \right)^{h_1+h_2+h_3+1} \left[ {(h_1-h_2)^2\over h_3}+{(h_2-h_3)^2\over h_1}+{(h_3-h_1)^2\over h_2} \right],~~~~~~~~~~~~~~~~~~~~~~~~~~~
\fe
\ie\nonumber
A_2& = {2^{-{c\over 2}-9}\left({16\over 27}\right)^{h_1+h_2+h_3+2}\over h_1(c+2h_1(c+8h_1-5))h_2(c+2h_2(c+8h_2-5))h_3(c+2h_3(c+8h_3-5))}\times~~~~~~~~~~~~
\fe
{\fontsize{1}{1}
$$\renewcommand*{\arraystretch}{.2} \begin{array}{l}
\bigg\{ 2048 \bigg[16 \left(c+8 h_3\right) h_2^3+2 \left(128 h_3^2+24 (c-5) h_3+c (c+3)\right) h_2^2+\left(128 h_3^3+48 (c-5) h_3^2+4 \left(c^2-6 c+25\right) h_3+c (3 c-10)\right) h_2+c \left(h_3+1\right) \left(16 h_3^2+2 (c-5) h_3+c\right)\bigg] h_1^7 +256 \bigg[-512 \left(c+8 h_3\right) h_2^4-16 \left(256 h_3^2+8 (7 c-23) h_3+c (3 c+1)\right) h_2^3+2 \left(-2048 h_3^3-128 (3 c+1) h_3^2-8 \left(c^2+66 c-195\right) h_3+c \left(c^2-34 c-63\right)\right) h_2^2 +\left(-4096 h_3^4-128 (7 c-23) h_3^3-16 \left(c^2+66 c-195\right) h_3^2+4 \left(c^3-43 c^2+247 c-525\right) h_3+c \left(3 c^2-73 c+210\right)\right) h_2+c \left(-512 h_3^4-16 (3 c+1) h_3^3+2 \left(c^2-34 c-63\right) h_3^2+\left(3 c^2-73 c+210\right) h_3+(c-21) c\right)\bigg] h_1^6 -128 \bigg[-1536 \left(c+8 h_3\right) h_2^5-64 \left(64 h_3^2+16 (c-2) h_3+(c-1) c\right) h_2^4+16 \left(-1024 h_3^3-32 (3 c+20) h_3^2+4 \left(3 c^2-96 c+227\right) h_3+c \left(c^2-15 c-49\right)\right) h_2^3+2 \big(-2048 h_3^4-256 (3 c+20) h_3^3-64 (33 c-103) h_3^2+8 \left(c^3-27 c^2+42 c-70\right) h_3+c \left(7 c^2-103 c+132\right)\big) h_2^2 \\ 
+\big(-12288 h_3^5-1024 (c-2) h_3^4+64 \left(3 c^2-96 c+227\right) h_3^3+16 \left(c^3-27 c^2+42 c-70\right) h_3^2+4 \left(7 c^3-186 c^2+779 c-900\right) h_3 +3 c \left(3 c^2-62 c+120\right)\big) h_2+c \left(-1536 h_3^5-64 (c-1) h_3^4+16 \left(c^2-15 c-49\right) h_3^3+2 \left(7 c^2-103 c+132\right) h_3^2+3 \left(3 c^2-62 c+120\right) h_3+3 (c-12) c\right)\bigg] h_1^5 +256 \bigg[-512 \left(c+8 h_3\right) h_2^6+32 \left(64 h_3^2+16 (c-2) h_3+(c-1) c\right) h_2^5+4 \left(-1536 h_3^3-64 (c+25) h_3^2+8 \left(5 c^2-106 c+241\right) h_3+c \left(3 c^2-38 c-93\right)\right) h_2^4+2 \big(-3072 h_3^4-256 (c+7) h_3^3+16 \left(2 c^2-57 c+75\right) h_3^2+2 \left(c^3-9 c^2-129 c+225\right) h_3+c \left(2 c^2-19 c-39\right)\big) h_2^3+\left(2048 h_3^5-256 (c+25) h_3^4+32 \left(2 c^2-57 c+75\right) h_3^3+8 \left(2 c^3-25 c^2-180 c+515\right) h_3^2+2 \left(12 c^3-235 c^2+821 c-950\right) h_3+c \left(5 c^2-123 c+200\right)\right) h_2^2+\big(-4096 h_3^6+512 (c-2) h_3^5+32 \left(5 c^2-106 c+241\right) h_3^4 +4 \left(c^3-9 c^2-129 c+225\right) h_3^3\\ 
+2 \left(12 c^3-235 c^2+821 c-950\right) h_3^2+2 \left(9 c^3-129 c^2+370 c-250\right) h_3+c \left(3 c^2-38 c+50\right)\big) h_2 +c \left(-512 h_3^6+32 (c-1) h_3^5+4 \left(3 c^2-38 c-93\right) h_3^4+\left(4 c^2-38 c-78\right) h_3^3+\left(5 c^2-123 c+200\right) h_3^2+\left(3 c^2-38 c+50\right) h_3-5 c\right)\bigg] h_1^4 -16 \bigg[-2048 \left(c+8 h_3\right) h_2^7+256 \left(256 h_3^2+8 (7 c-23) h_3+c (3 c+1)\right) h_2^6+128 \left(-1024 h_3^3-32 (3 c+20) h_3^2+4 \left(3 c^2-96 c+227\right) h_3+c \left(c^2-15 c-49\right)\right) h_2^5-32 \big(-3072 h_3^4-256 (c+7) h_3^3+16 \left(2 c^2-57 c+75\right) h_3^2 +2 \left(c^3-9 c^2-129 c+225\right) h_3+c \left(2 c^2-19 c-39\right)\big) h_2^4+16 \left(-8192 h_3^5+512 (c+7) h_3^4+48 \left(6 c^2-68 c-347\right) h_3^3+2 \left(6 c^3-31 c^2-1495 c+5600\right) h_3^2+\left(22 c^3-469 c^2+228 c-600\right) h_3+c \left(3 c^2-102 c+40\right)\right) h_2^3+2 \big(32768 h_3^6-2048 (3 c+20) h_3^5-256 \left(2 c^2-57 c+75\right) h_3^4+16 \left(6 c^3-31 c^2-1495 c+5600\right) h_3^3-2 \left(2 c^3+365 c^2-6448 c+18225\right) h_3^2 \\ 
+\left(6 c^3-1419 c^2+7085 c-4000\right) h_3+4 c \left(3 c^2-40 c+100\right)\big) h_2^2 -\left(16384 h_3^7-2048 (7 c-23) h_3^6-512 \left(3 c^2-96 c+227\right) h_3^5+64 \left(c^3-9 c^2-129 c+225\right) h_3^4-16 \left(22 c^3-469 c^2+228 c-600\right) h_3^3-2 \left(6 c^3-1419 c^2+7085 c-4000\right) h_3^2+c \left(162 c^2+761 c-800\right) h_3+32 c^3\right) h_2 +8 c \left(-256 h_3^7+32 (3 c+1) h_3^6+16 \left(c^2-15 c-49\right) h_3^5+\left(-8 c^2+76 c+156\right) h_3^4+\left(6 c^2-204 c+80\right) h_3^3+\left(3 c^2-40 c+100\right) h_3^2-4 c^2 h_3-c^2\right)\bigg] h_1^3 +2 \bigg[2048 \left(128 h_3^2+24 (c-5) h_3+c (c+3)\right) h_2^7+256 \left(-2048 h_3^3-128 (3 c+1) h_3^2-8 \left(c^2+66 c-195\right) h_3+c \left(c^2-34 c-63\right)\right) h_2^6-128 \big(-2048 h_3^4-256 (3 c+20) h_3^3-64 (33 c-103) h_3^2+8 \left(c^3-27 c^2+42 c-70\right) h_3+c \left(7 c^2-103 c+132\right)\big) h_2^5 +128 \left(2048 h_3^5-256 (c+25) h_3^4+32 \left(2 c^2-57 c+75\right) h_3^3+8 \left(2 c^3-25 c^2-180 c+515\right) h_3^2+2 \left(12 c^3-235 c^2+821 c-950\right) h_3+c \left(5 c^2-123 c+200\right)\right) h_2^4\\ 
 -16 \left(32768 h_3^6-2048 (3 c+20) h_3^5-256 \left(2 c^2-57 c+75\right) h_3^4+16 \left(6 c^3-31 c^2-1495 c+5600\right) h_3^3-2 \left(2 c^3+365 c^2-6448 c+18225\right) h_3^2+\left(6 c^3-1419 c^2+7085 c-4000\right) h_3+4 c \left(3 c^2-40 c+100\right)\right) h_2^3 +2 \big(131072 h_3^7-16384 (3 c+1) h_3^6+4096 (33 c-103) h_3^5+512 \left(2 c^3-25 c^2-180 c+515\right) h_3^4+16 \left(2 c^3+365 c^2-6448 c+18225\right) h_3^3+6 \left(25 c^4-427 c^3-1605 c^2+17175 c-32000\right) h_3^2 +c \left(75 c^3-1078 c^2-997 c+12800\right) h_3+128 c^2 (5-2 c)\big) h_2^2+\big(49152 (c-5) h_3^7-2048 \left(c^2+66 c-195\right) h_3^6-1024 \left(c^3-27 c^2+42 c-70\right) h_3^5+256 \left(12 c^3-235 c^2+821 c-950\right) h_3^4-16 \left(6 c^3-1419 c^2+7085 c-4000\right) h_3^3+2 c \left(75 c^3-1078 c^2-997 c+12800\right) h_3^2+5 c^2 \left(15 c^2-15 c-512\right) h_3-64 c^3\big) h_2 +64 c h_3 \big(32 (c+3) h_3^6+4 \left(c^2-34 c-63\right) h_3^5-2 \left(7 c^2-103 c+132\right) h_3^4+2 \left(5 c^2-123 c+200\right) h_3^3+\left(-3 c^2+40 c-100\right) h_3^2
\\ 
 +4 c (5-2 c) h_3-c^2\big)\bigg] h_1^2 +\bigg[2048 \left(128 h_3^3+48 (c-5) h_3^2+4 \left(c^2-6 c+25\right) h_3+c (3 c-10)\right) h_2^7+256 \left(-4096 h_3^4-128 (7 c-23) h_3^3-16 \left(c^2+66 c-195\right) h_3^2+4 \left(c^3-43 c^2+247 c-525\right) h_3+c \left(3 c^2-73 c+210\right)\right) h_2^6 -128 \left(-12288 h_3^5-1024 (c-2) h_3^4+64 \left(3 c^2-96 c+227\right) h_3^3+16 \left(c^3-27 c^2+42 c-70\right) h_3^2+4 \left(7 c^3-186 c^2+779 c-900\right) h_3+3 c \left(3 c^2-62 c+120\right)\right) h_2^5+256 \big(-4096 h_3^6+512 (c-2) h_3^5 +32 \left(5 c^2-106 c+241\right) h_3^4+4 \left(c^3-9 c^2-129 c+225\right) h_3^3 +2 \left(12 c^3-235 c^2+821 c-950\right) h_3^2+2 \left(9 c^3-129 c^2+370 c-250\right) h_3+c \left(3 c^2-38 c+50\right)\big) h_2^4+16 \big(16384 h_3^7-2048 (7 c-23) h_3^6-512 \left(3 c^2-96 c+227\right) h_3^5+64 \left(c^3-9 c^2-129 c+225\right) h_3^4-16 \left(22 c^3-469 c^2+228 c-600\right) h_3^3-2 \left(6 c^3-1419 c^2+7085 c-4000\right) h_3^2+c \left(162 c^2+761 c-800\right) h_3+32 c^3\big) h_2^3+2 \big(49152 (c-5) h_3^7
\\ 
 -2048 \left(c^2+66 c-195\right) h_3^6-1024 \left(c^3-27 c^2+42 c-70\right) h_3^5+256 \left(12 c^3-235 c^2+821 c-950\right) h_3^4-16 \left(6 c^3-1419 c^2+7085 c-4000\right) h_3^3+2 c \left(75 c^3-1078 c^2-997 c+12800\right) h_3^2+5 c^2 \left(15 c^2-15 c-512\right) h_3 -64 c^3\big) h_2^2+h_3 \big(8192 \left(c^2-6 c+25\right) h_3^6+1024 \left(c^3-43 c^2+247 c-525\right) h_3^5-512 \left(7 c^3-186 c^2+779 c-900\right) h_3^4+512 \left(9 c^3-129 c^2+370 c-250\right) h_3^3+16 c \left(162 c^2+761 c-800\right) h_3^2+10 c^2 \left(15 c^2-15 c-512\right) h_3 +3 c^3 (25 c+256)\big) h_2+128 c h_3^2 \left(16 (3 c-10) h_3^5+2 \left(3 c^2-73 c+210\right) h_3^4-3 \left(3 c^2-62 c+120\right) h_3^3+2 \left(3 c^2-38 c+50\right) h_3^2+4 c^2 h_3-c^2\right)\bigg] h_1+128 c \left(16 h_2^2+2 (c-5) h_2+c\right) \left(h_2-h_3\right){}^2 \left(16 h_3^2+2 (c-5) h_3+c\right) \left(\left(h_3+1\right) h_2^3-2 \left(h_3^2+1\right) h_2^2+\left(h_3^3+1\right) h_2+\left(h_3-1\right){}^2 h_3\right)\bigg\}.
\end{array}
$$
}
We also record here the first few coefficients $A_n^0$ in the $q$-expansion of the vacuum block in the pillow frame analogous to (\ref{pillowexpansion}), which, as already emphasized, differ from the $h_i\to 0$ limit of (\ref{acoeffs}),
\ie
&A^0_0 = 2^{-{c\over 2}},~~~ A^0_1 = 0,~~~ A^0_2 = 2^{-{c\over 2}-1}{25c\over 243},~~~ A^0_3 = 0,\\
&A^0_4 = 2^{-{c\over 2}-3}{1875 c^2 + 83110 c + 524288\over 177147},~~~ A^0_5 = {2^{-{c\over 2}+21}\over 4782969}, \\
&A^0_6 = 2^{-{c\over 2}-4}{140625c^3 + 18699750  c^2 + 349131040 c + 2969567232 + 2147483648c^{-1}\over 387420489}.
\fe
Importantly, all of the coefficients $A_n(h_1, h_2, h_3)$ are non-negative, as they can be interpreted as inner products of level $n$ descendant states created by pairs of twist-anti-twist fields on two corners of the pillow, similarly to the sphere 4-point block analyzed in \cite{Maldacena:2015iua}. Indeed, we have explicitly verified the positivity of $A_n(h_1, h_2, h_3)$ with $c>1$ and $h_i>0$, for $n\leq 5$.

\section{The genus two modular crossing equation}
\label{crossing}

\subsection{Some preliminary analysis}

Now we consider the genus two modular crossing equation restricted to the Renyi surface, as given by (\ref{modc}). Some crude but rigorous constraints on the structure constants in unitary CFTs can be deduced even without appealing to the details of the $z$-expansion of the genus two conformal block. First, let us write the twist field 4-point function (\ref{twist4}) in the pillow coordinates,
\ie\label{pils}
\left\langle \sigma_3(0)\bar\sigma_3(z,\bar z)\sigma_3(1)\bar \sigma_3(\infty) \right\rangle &= \left| (z(1-z))^{{c\over 8} - 2 h_\sigma} \theta_3(\tau)^{{3c\over 2} - 16 h_\sigma} q^{-{c\over 8}} \right|^2 
\\
&~~~\times \sum_{i,j,k}\sum_{n,m=0}^\infty  C_{ijk}^2 A_n(h_i, h_j, h_k) A_m(\tilde h_i, \tilde h_j,\tilde h_k) q^{h_i+h_j+h_k + n} \bar q^{\tilde h_i+\tilde h_j+\tilde h_k+ m}
\\
&= \left| (z(1-z))^{{c\over 8} - 2 h_\sigma} \theta_3(\tau)^{{3c\over 2} - 16 h_\sigma} q^{-{c\over 8}} \right|^2 \sum_{(h,\tilde h)\in {\cal J}} \widetilde C^2_{h,\tilde h} q^h \bar q^{\tilde h}.
\fe
In the last line, we simply grouped terms of the same powers of $q$ and $\bar q$ together in the sum. The index set ${\cal J}$ is by construction the union of $(\sum_{i=1}^3 h_i+\mathbb{Z}_{\geq 0}, \sum_{i=1}^3 \tilde h_i+\mathbb{Z}_{\geq 0})$
for all triples of conformal weights $\{(h_i, \tilde h_i), i=1,2,3\}$ that appear in nonzero structure constants, including the case where one of the primaries is the identity and the structure constant reduces to the two-point function coefficient.
It follows from the non-negativity of the coefficients $A_n$ that $\widetilde C_{h,\tilde h}^2$ are non-negative quantities in a unitary CFT.

Let us now apply (\ref{pils}) to a unitary noncompact CFT, where the $SL(2)$-invariant vacuum is absent and the identity is not included in the spectrum of ($\delta$-function) normalizable operators. $\widetilde C_{h,\tilde h}^2$ now only receives contributions from the structure constants of nontrivial primaries. Applying first order derivatives in $z$ and $\bar z$ to the crossing equation, and evaluating at $z=\bar z ={1\over 2}$, we have
\ie\label{firstcross}
\sum_{(h,\tilde h)\in {\cal J}'} \widetilde C^2_{h,\tilde h} \left.\partial_z\right|_{z={1\over 2}}\left[ (z(1-z))^{{c\over 4} - 4 h_\sigma} \theta_3(\tau)^{{3c} - 32 h_\sigma} q^{h+\tilde h -{c\over 4}} \right] = 0.
\fe
In the above equation, the factor multiplying $\widetilde C_{h,\tilde h}^2$ is negative for $\Delta \equiv h +\tilde h$ below a certain ``critical dimension" $\Delta_{crit}$ and positive for $\Delta>\Delta_{crit}$. It follows immediately that there must be a nonzero $\widetilde C_{h,\tilde h}^2$ for $\Delta<\Delta_{crit}$, i.e. there must be a triple of primaries with nonzero structure constant, whose total scaling dimension is less than $\Delta_{crit}$, in any unitary noncompact CFT of central charge $c$. The value of (or rather, an upper bound on) the critical dimension is easily computed from (\ref{firstcross}) to be
\ie\label{delc}
\Delta_{crit} = \left( {1\over 4} - {3\over 4\pi} \right) c + {8\over \pi} h_\sigma = {9\pi + 5\over 36\pi}c \approx 0.29421 c. 
\fe
As a consistency check, the Liouville CFT of central charge $c$ has nonzero structure constants for triples of primaries of total scaling dimension above the threshold ${{c-1}\over 4}$, which is indeed less than (\ref{delc}). 

Although rigorous, the bound (\ref{delc}) is quite crude. To deduce similar results in compact CFTs, it will be important to distinguish the contributions of Virasoro descendants from those of the primaries in (\ref{pils}). We will refine our analysis in the next subsection by computing the $z$ or $q$-expansion of the genus two conformal block to higher orders.

\subsection{Critical surfaces}

As is standard in the numerical bootstrap \cite{Rattazzi:2008pe, Rychkov:2009ij, Simmons-Duffin:2016gjk}, we can turn the genus two modular crossing equation (\ref{modc}) into linear equations for $C_{ijk}^2$ by acting on it with the linear functional
\ie
\A = \left. \sum_{n+m={\rm odd}} a_{n,m} \partial_z^n \partial_{\bar z}^m \right|_{z=\bar z={1\over 2}},
\fe
where $a_{n,m}$ are a set of real coefficients, and obtain constraints on the structure constants of the general form
\ie\label{conseq}
\sum_{i,j,k\in{\cal I}} C_{ijk}^2 F_c^{\A}(h_i, h_j, h_k; \tilde h_i, \tilde h_j, \tilde h_k) = 0,
\fe
where $F_c^\A$ is a function of a triple of left and right conformal weights. For typical choices of the linear functional $\A$, $F_c^\A$ will be negative on a domain $D$ in the space of triples of conformal weights, and positive on the complement of the closure of $D$. A critical surface $S$ is defined to be the boundary of $D$ where $F_c^\A$ vanishes. With an appropriate choice of sign in $\A$, the domain $D$ consists of triples of low lying weights (we will see that the critical surface is often compact), and the equation (\ref{conseq}) implies that the structure constants outside of $D$ are bounded by those that lie within $D$.

\begin{figure}[h!]
\centering
\subfloat{\includegraphics[width=.33\textwidth]{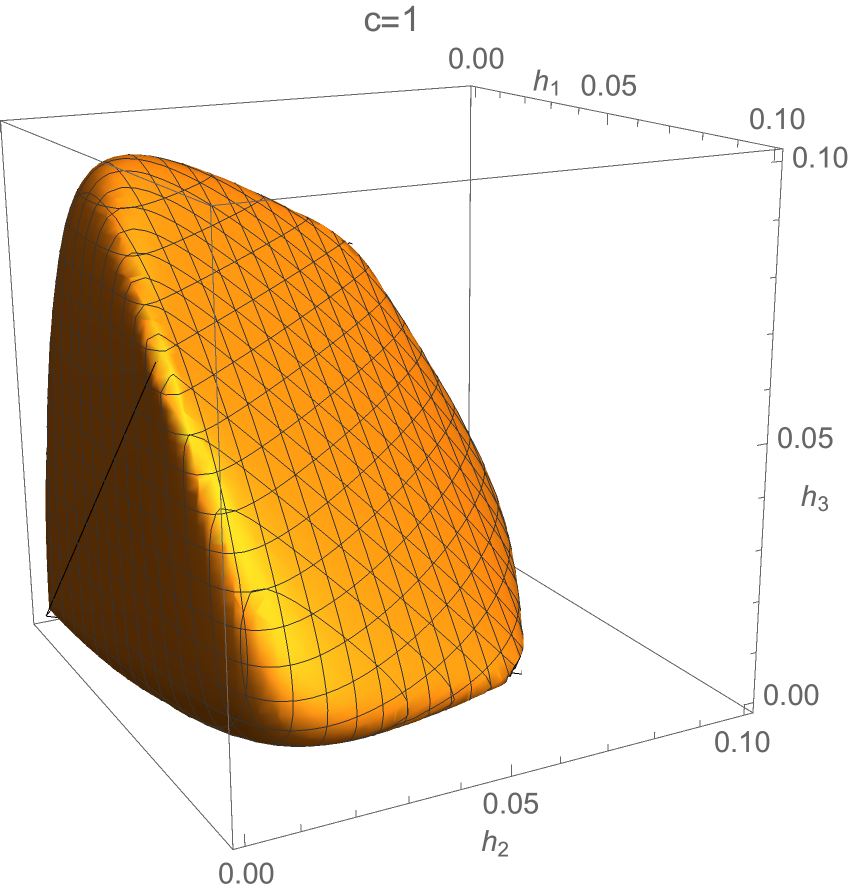}}
\subfloat{\includegraphics[width=.33\textwidth]{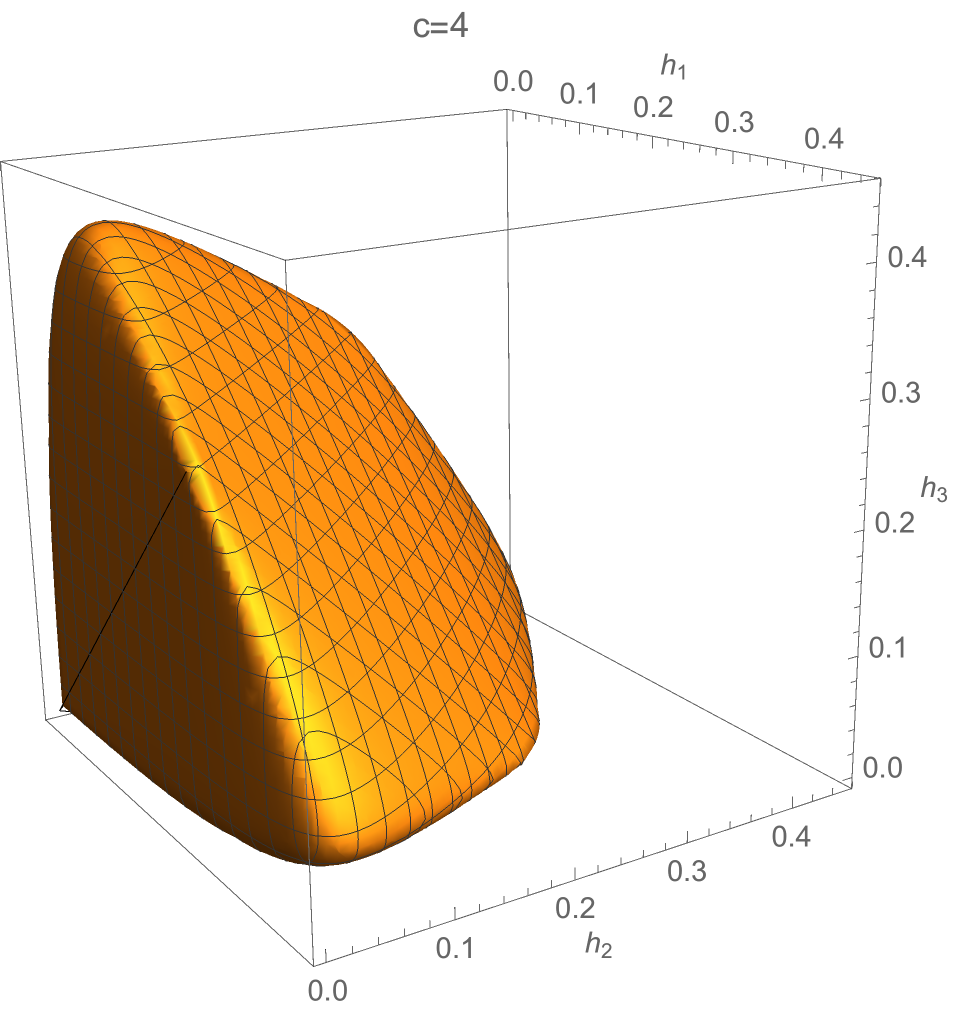}}
\subfloat{\includegraphics[width=.33\textwidth]{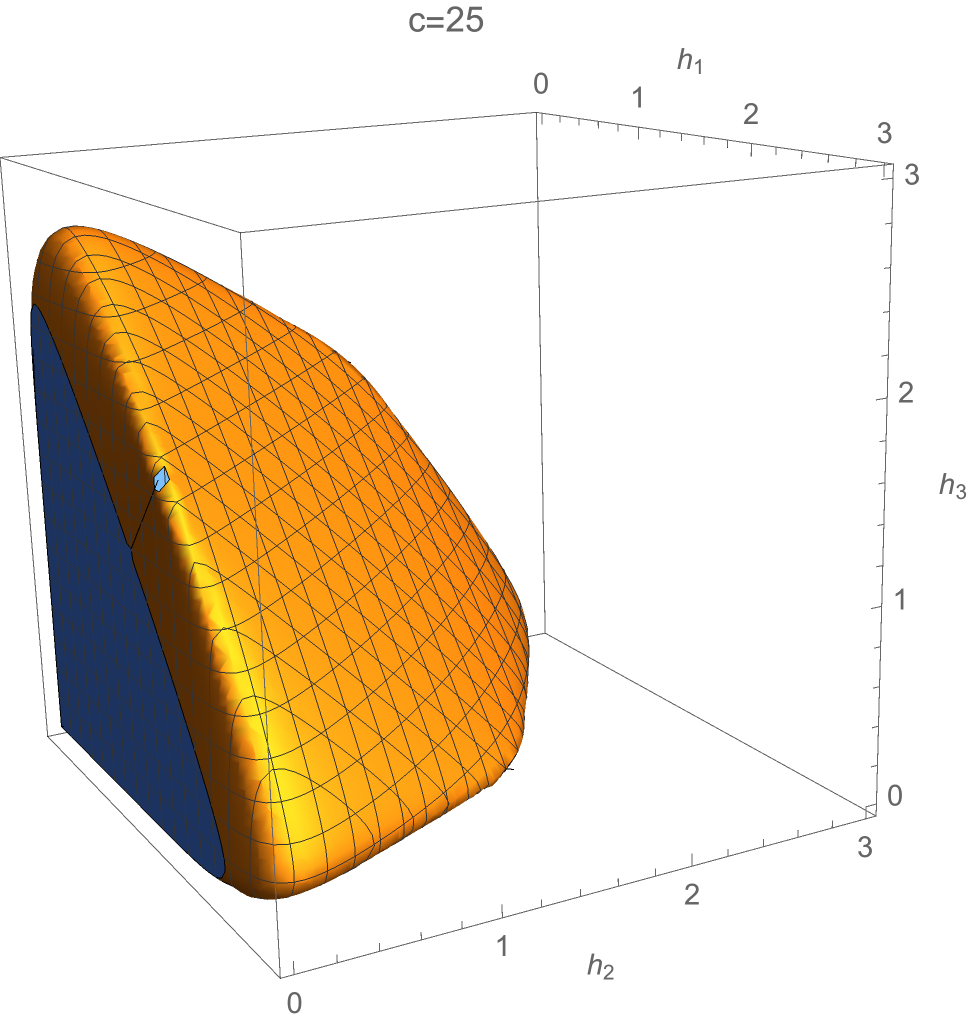}}
\\
\subfloat{\includegraphics[width=.33\textwidth]{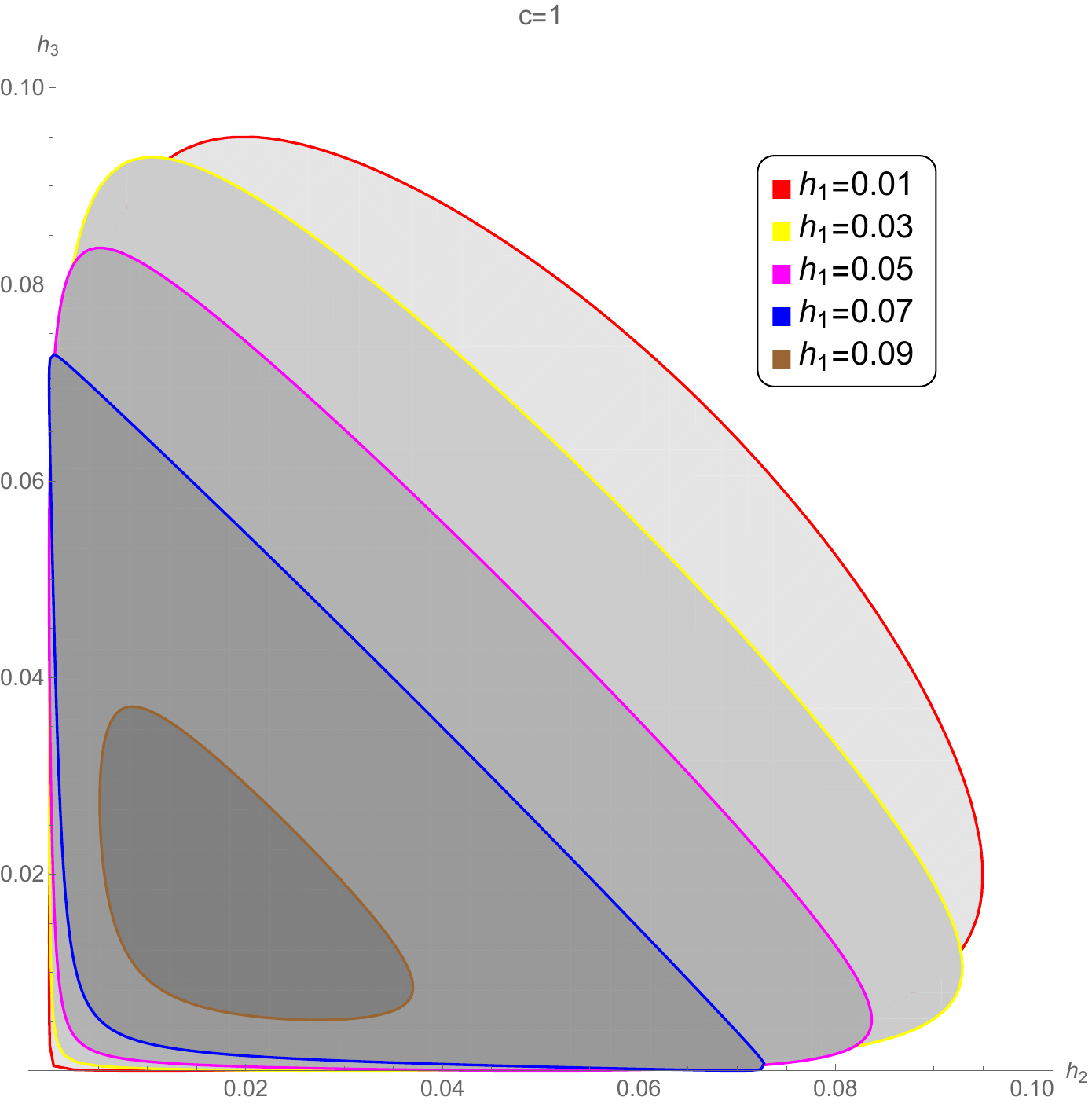}}
\subfloat{\includegraphics[width=.33\textwidth]{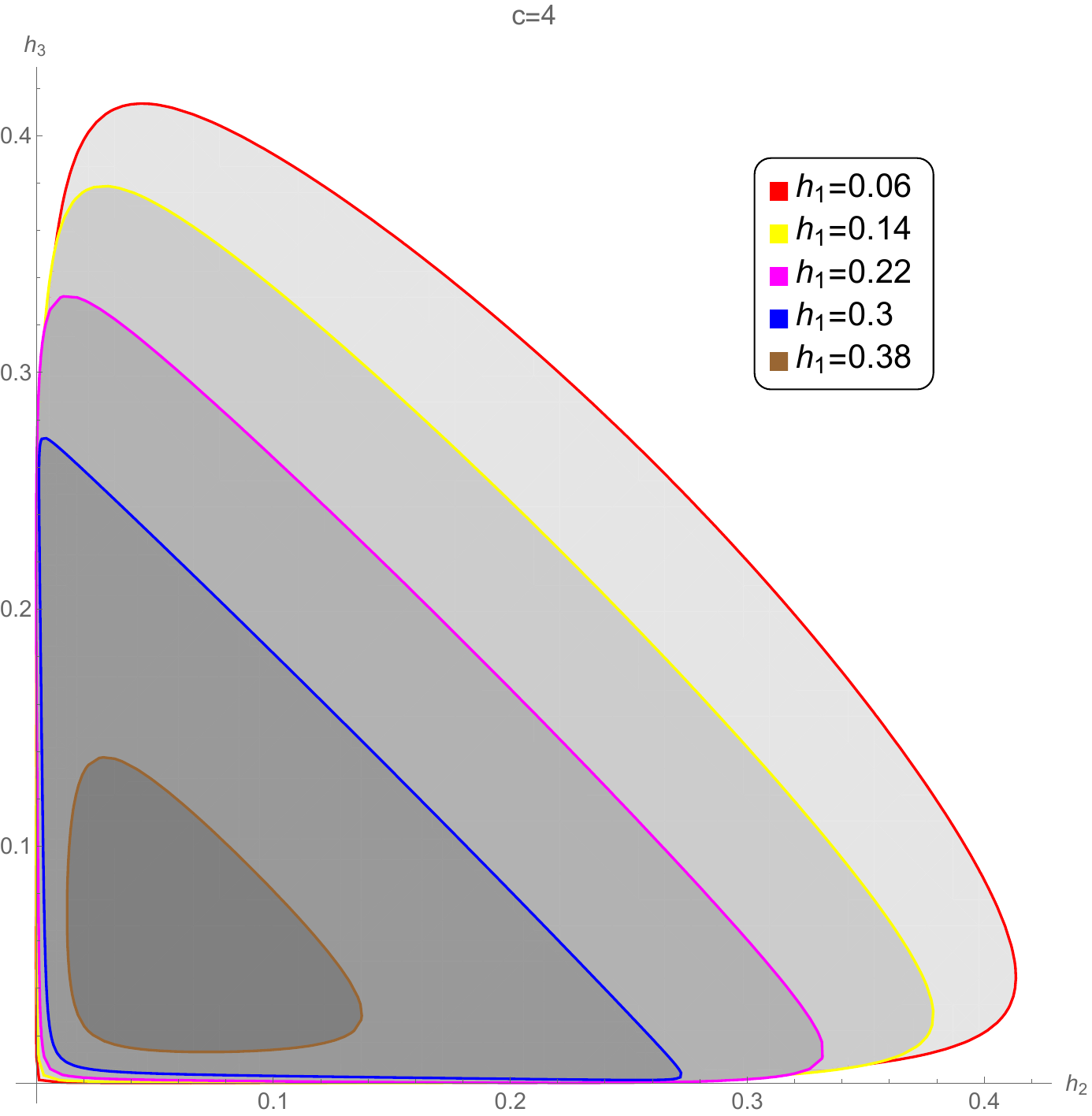}}
\subfloat{\includegraphics[width=.33\textwidth]{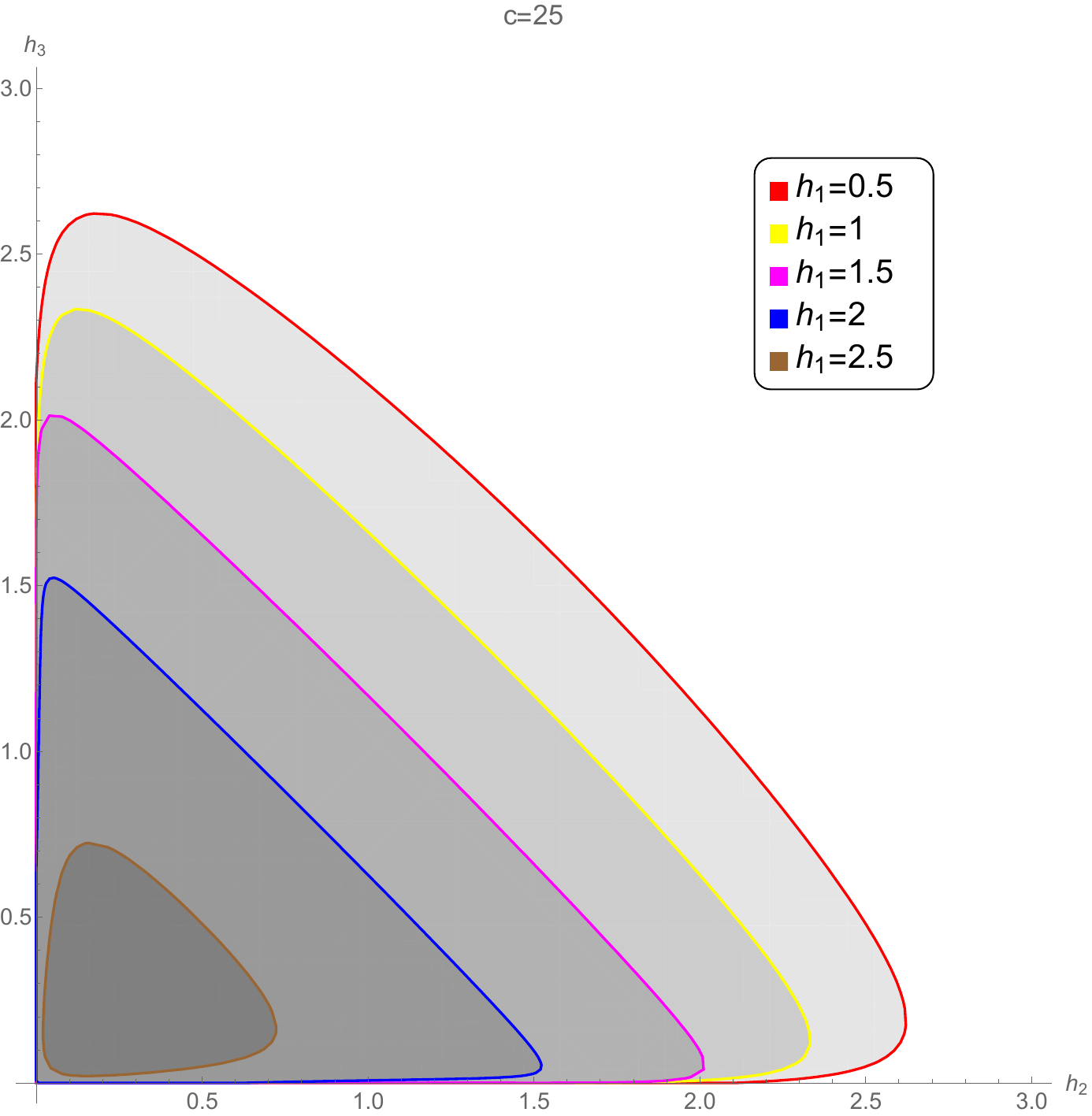}}
\caption{{\bf Top}: Three-dimensional plots of the domain $D^{(3)}_h$ for $c=1,4,25$. {\bf Bottom}: Plots of the cross-sections of these domains for various values of $h_1$. The structure constants of primaries with twists $(\tau_1,\tau_2,\tau_3) = (2h_1,2h_2,2h_3)$ outside these critical domains are bounded by those whose twists lie within the domains.}\label{fig:CriticalTwist}
\end{figure}

Clearly, the critical surface $S$ depends on the choice of $\A$. It is of interest to find critical surfaces that bound a domain $D$ that is as ``small" as possible, so that we can bound as many structure constants as possible based on the knowledge of a small set of structure constants of low dimension operators in any unitary CFT. Here we will consider the simplest nontrivial linear functional $\A$ which involves only first order derivatives in $z$ or in $\bar z$. In this case, the critical surface is the locus
\ie{}
&a_{1,0} W_c(h_1, h_2, h_3) + a_{0,1}W_c(\tilde h_1,\tilde h_2,\tilde h_3) = 0,
\fe
where $W_c(h_1, h_2, h_3)\equiv \left.\partial_z \log {\cal F}_c(h_1,h_2,h_3|z)\right|_{z={1\over 2}}$. For instance, we can choose $a_{0,1}=0$, and the critical surface $W_c(h_1, h_2, h_3)=0$ bounds a {\it compact} domain $D_h$ in $\mathbb{R}^3_{\geq 0}$ parameterized by the holomorphic weights $(h_1, h_2, h_3)$, and bound structure constants of triples of primaries of higher twists by those of lower twists. 
From (\ref{pillowexpansion}), we have
\ie\label{eq:Wc}
W_c(h_1,h_2,h_3)={\pi^2\over K({1\over 2})^2}\left[h_1+h_2+h_3 -\left({1\over 8} + {5\over 72\pi} \right) c+{\sum_{n=1}^{\infty}nA_{n}(h_1, h_2, h_3) e^{-n\pi}\over \sum_{n=0}^{\infty}A_n(h_1, h_2, h_3) e^{-n\pi}} \right].
\fe
The last term in the bracket is always positive (assuming $c>1$ and $h_i>0$), thus the domain $W_c<0$ lies within the region $h_1+h_2+h_3 < \left({1\over 8} + {5\over 72\pi} \right) c$ and is compact. This is what we have seen in the previous subsection. 

\begin{figure}[h!]
\centering
\subfloat{\includegraphics[width=1\textwidth]{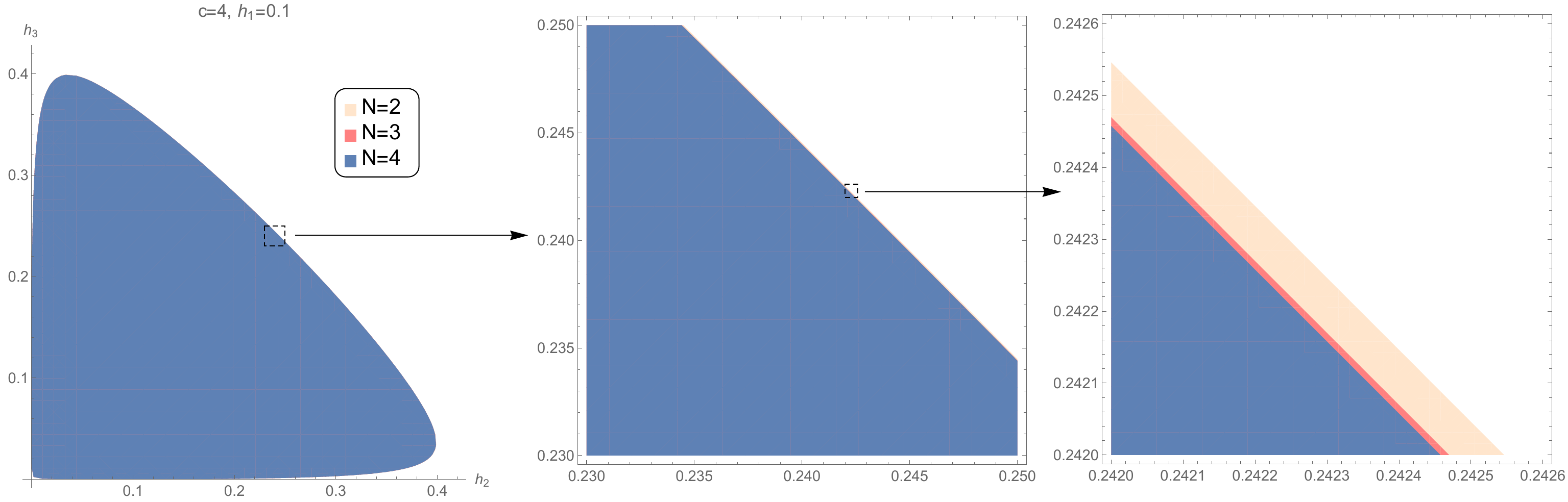}}
\caption{A slice of the $c=4$ critical domain $D_h^{(N)}$, which converges quickly with the truncation order $N$ of the $q$-expansion.}\label{fig:convergence}
\end{figure}

For numerical evaluation we may work with the truncated version
\ie{}
W_c^{(N)}(h_1,h_2,h_3)={\pi^2\over K({1\over 2})^2}\left[h_1+h_2+h_3 -\left({1\over 8} + {5\over 72\pi} \right) c+{\sum_{n=1}^{N}nA_{n}(h_1, h_2, h_3) e^{-n\pi}\over \sum_{n=0}^{N}A_n(h_1, h_2, h_3) e^{-n\pi}} \right].
\fe
The domain $D_h^{(N)} = \{(h_1, h_2, h_3)\in \mathbb{R}^3_{\geq 0}: W_c^{(N)}(h_1, h_2, h_3)<0\}$ becomes smaller with increasing $N$ (and of course, converges to $D_h$ in the $N\to \infty$ limit). In Figure \ref{fig:CriticalTwist} we plot some examples of the critical domain $D_h^{(N)}$ with $N=3$, for central charges $c=1$, 4, and 25. The location of the critical surface converges rather quickly with the $q$-expansion order: an example of a slice of the critical domain in the $c=4$ case is shown in Figure \ref{fig:convergence}.

\begin{figure}[h!]
\centering
\subfloat{\includegraphics[width=.49\textwidth]{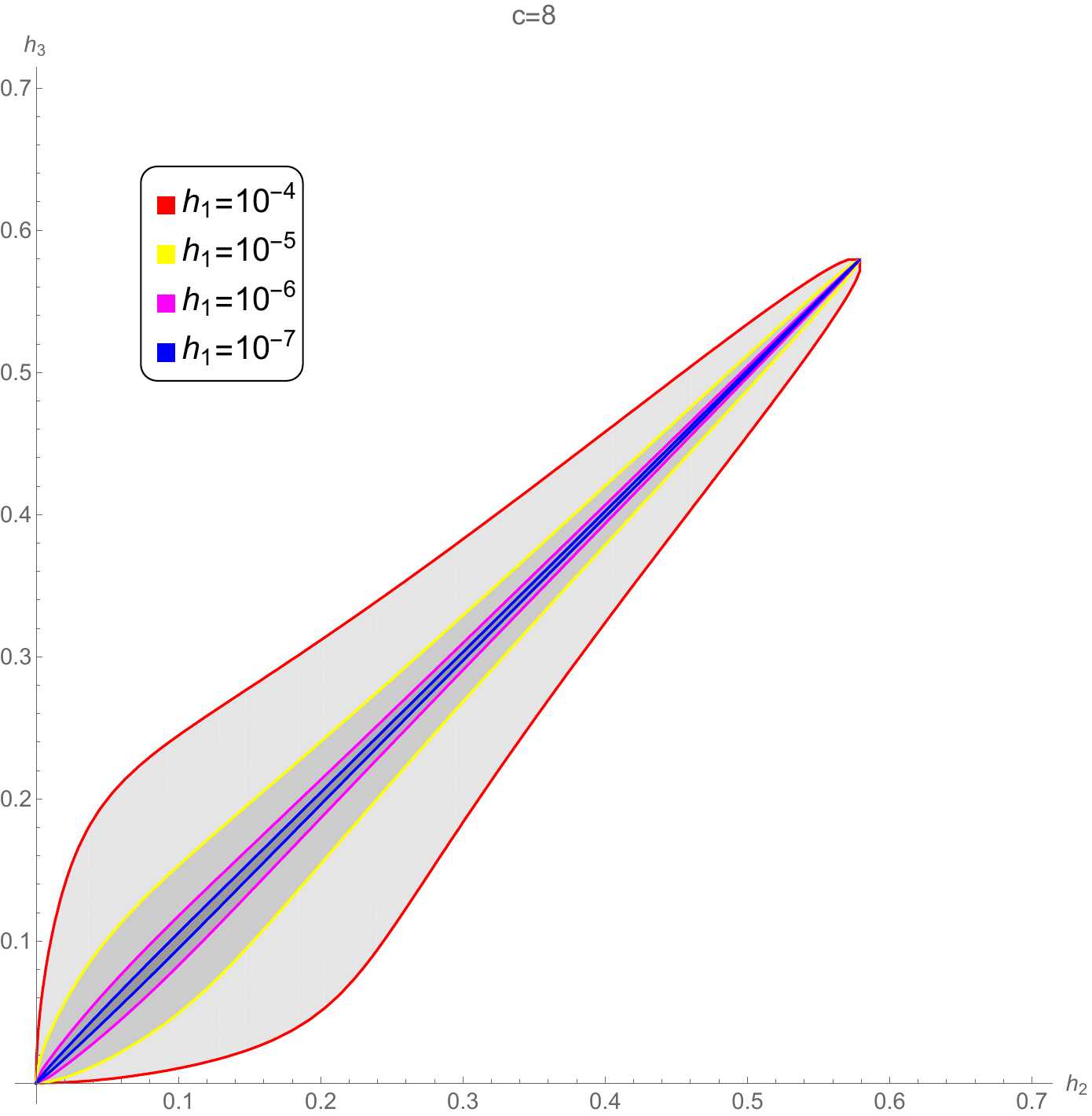}}
\subfloat{\includegraphics[width=.49\textwidth]{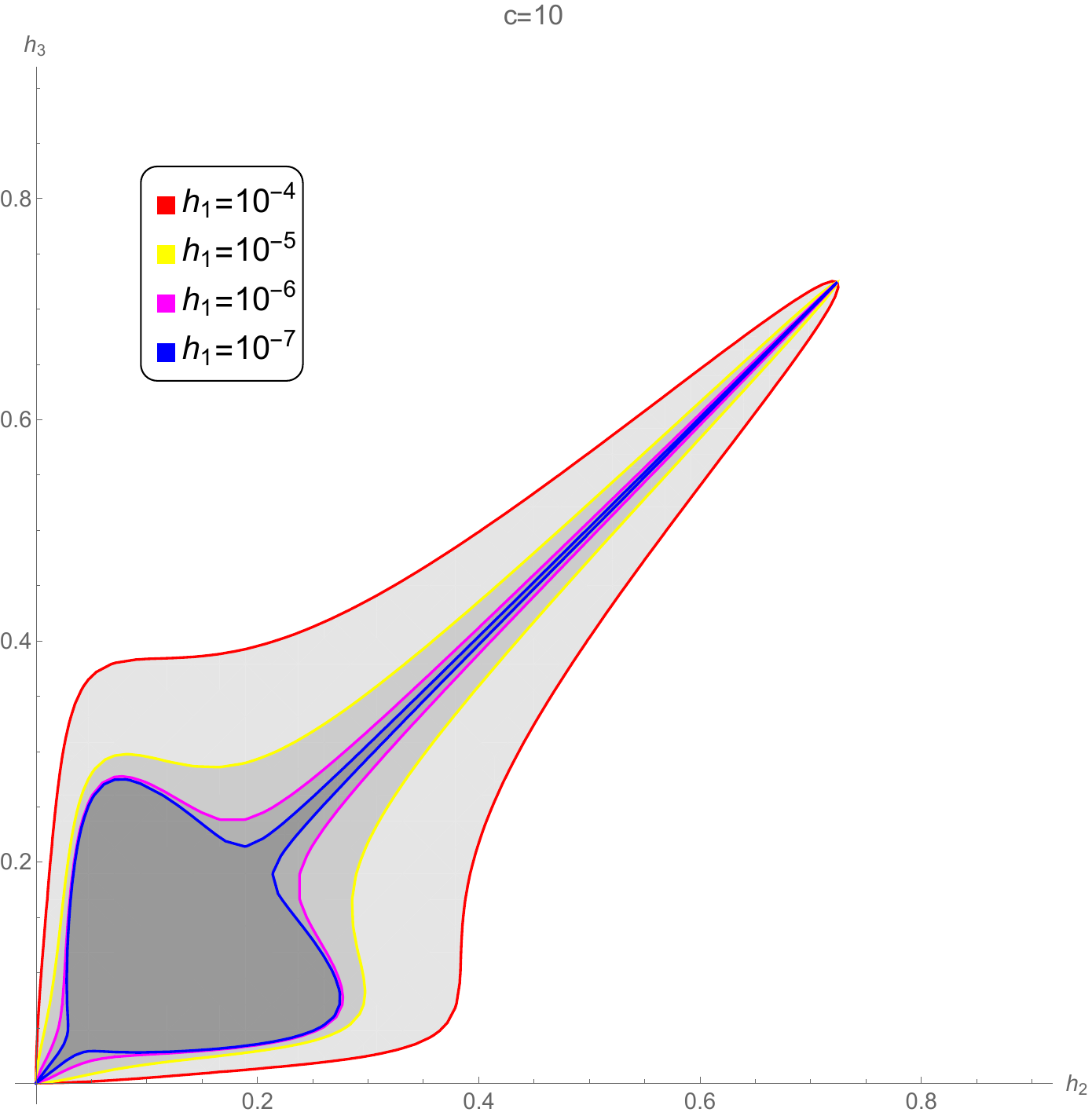}}
\caption{Slices of the critical domain $D_h^{(3)}$ for $c=8$ and $c=10$ at small values of $h_1$. For $c=8$, the critical domain $D_h$ intersects the $h_1=0$ plane only along a segment of the diagonal line $h_2=h_3$. This is not the case for $c=10$.}\label{fig:SmallDiagonal}
\end{figure}
 
In particular, in the limit $h_1\to 0$, with $h_2, h_3$ fixed at generic positive values, the coefficients $A_n$ diverge like $h_1^{-1} P_n(h_2, h_3)$, where $P_n$ is a rational function of $h_2, h_3$ that vanishes quadratically along $h_2=h_3$ ($>0$). For $h_2\not= h_3>0$, for instance, we have $\lim_{h_1\to 0} W_c^{(1)}(h_1, h_2, h_3) = {\pi^2\over K({1\over 2})^2}\left[ h_2+h_3 - \left({1\over 8} + {5\over 72\pi} \right) c + 1 \right]$, which is always positive for $c<6.79787$. A slightly more intricate analysis of  $\lim_{h_1\to 0} W_c^{(2)}$ shows that it is positive for $c<9.31751$. Consequently, for this range of the central charge $c$, the domain $D^{(2)}_h$ (and thereby $D_h$) meets the $h_1=0$ plane along a segment of the line $h_2=h_3$ only. This is demonstrated in Figure \ref{fig:SmallDiagonal}.

\begin{figure}[h!]
\centering
\subfloat{\includegraphics[width=.33\textwidth]{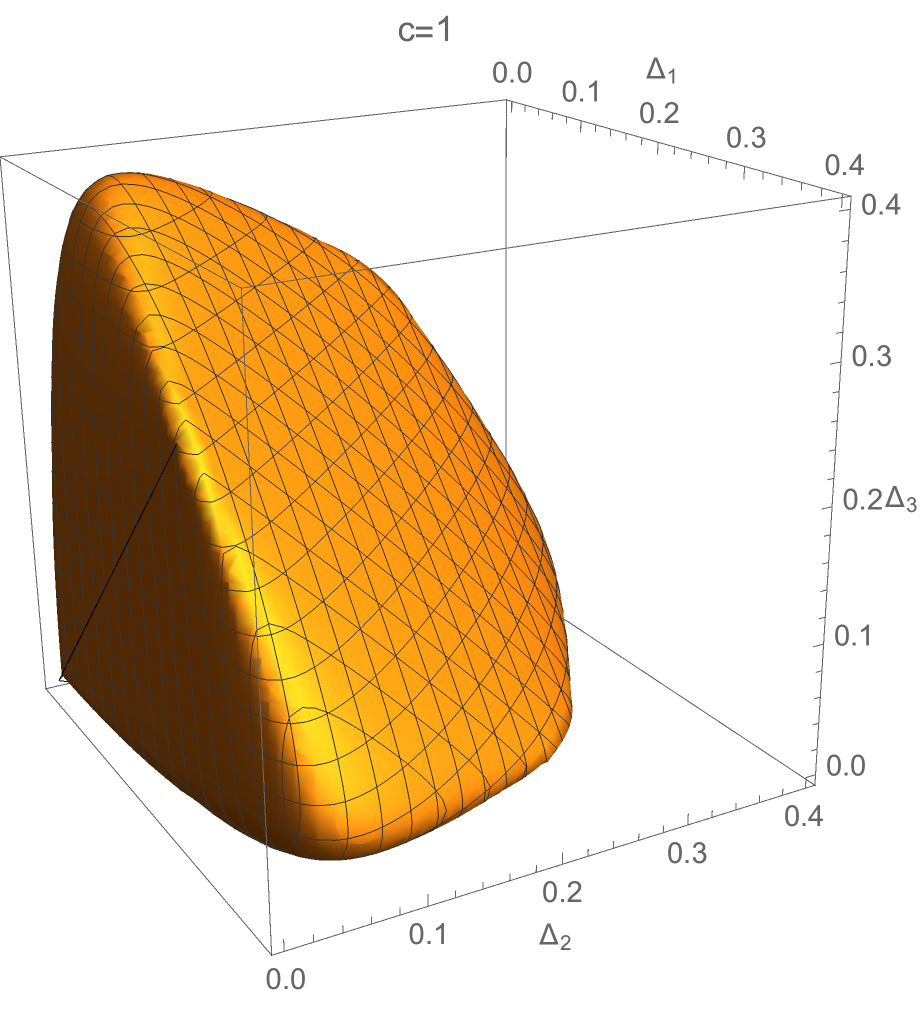}}
\subfloat{\includegraphics[width=.33\textwidth]{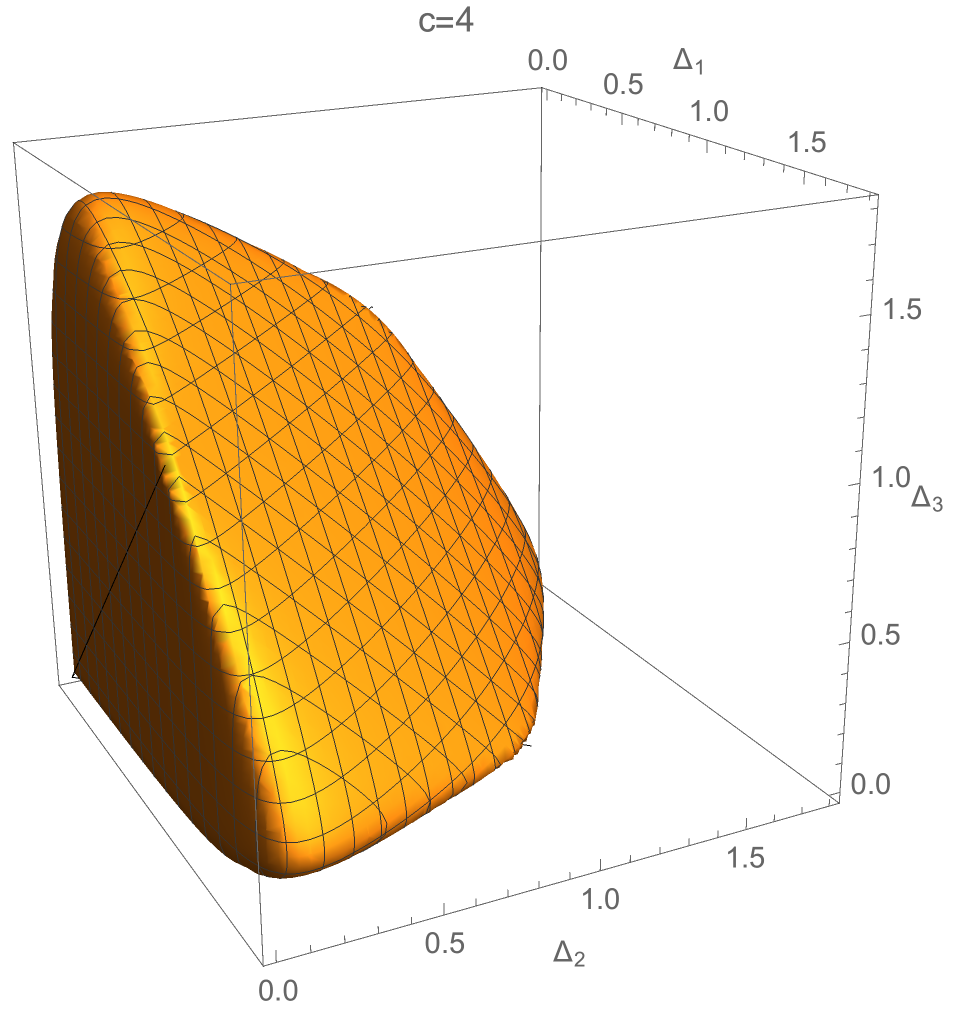}}
\subfloat{\includegraphics[width=.33\textwidth]{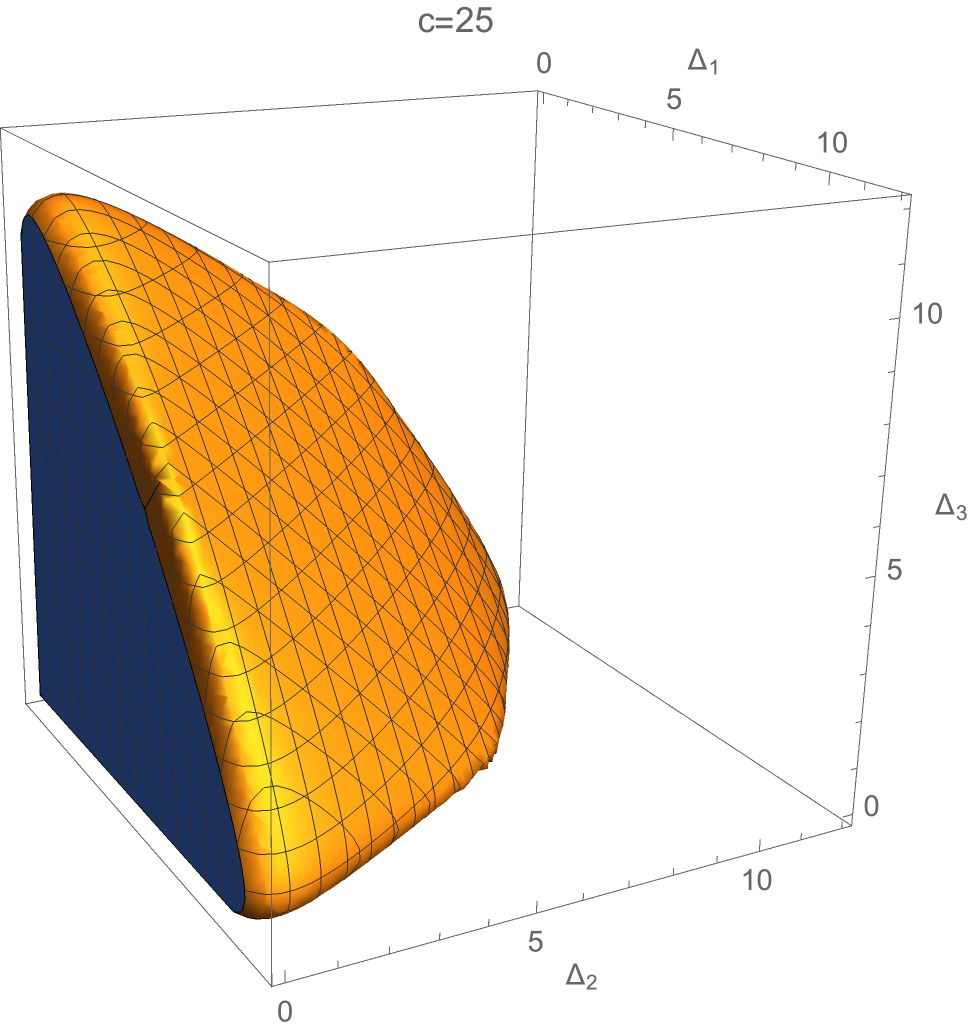}}
\\
\subfloat{\includegraphics[width=.33\textwidth]{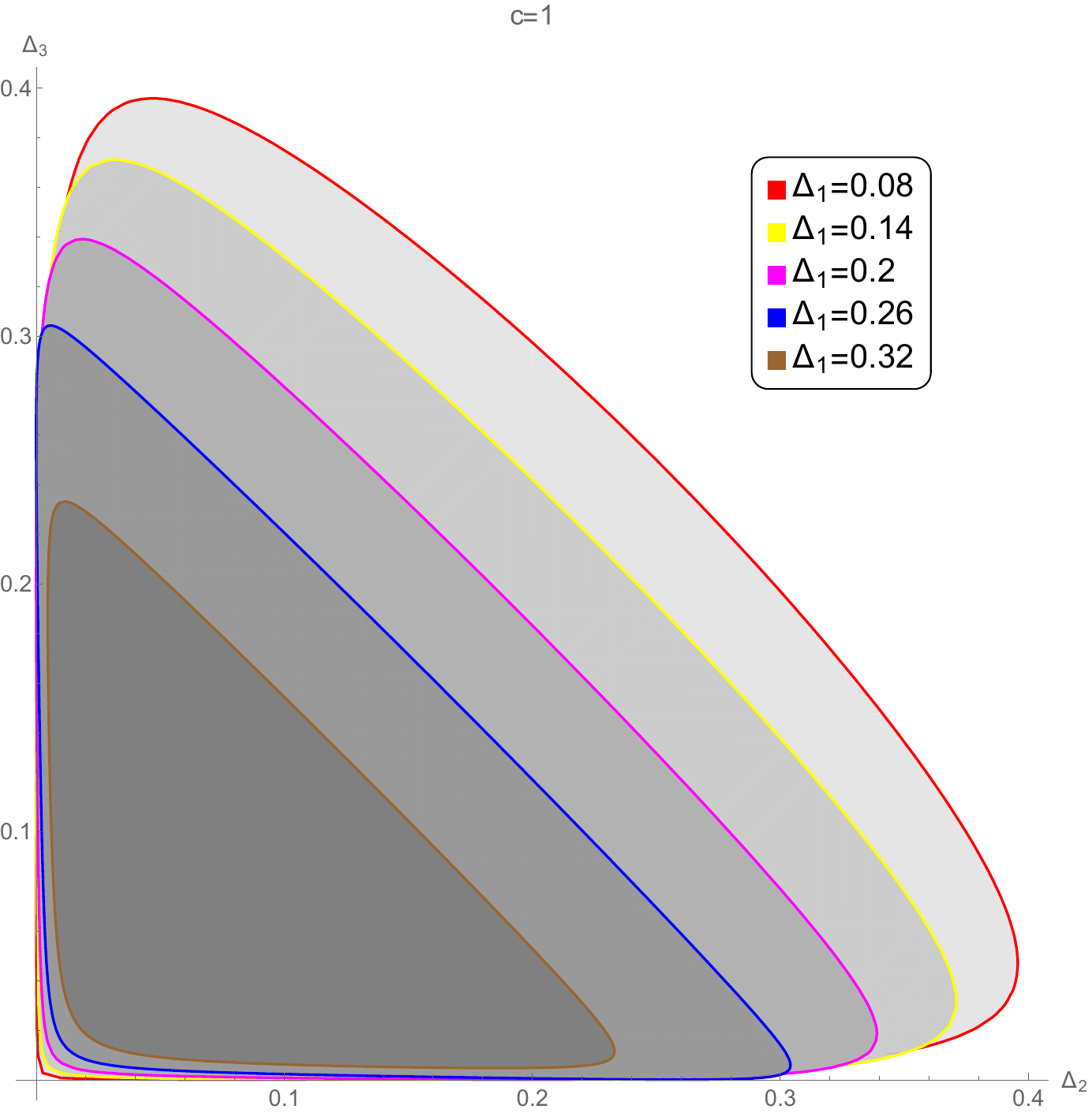}}
\subfloat{\includegraphics[width=.33\textwidth]{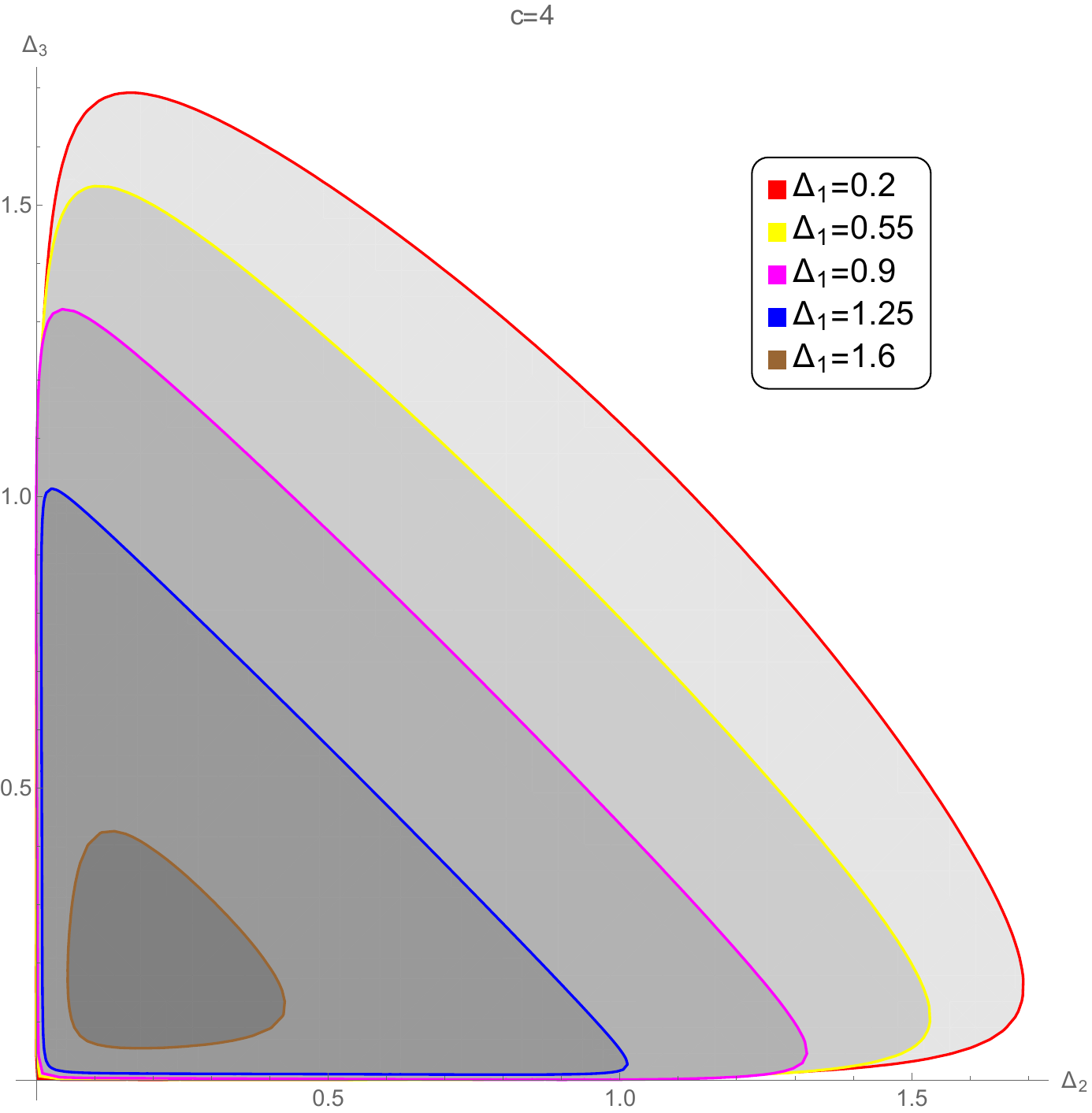}}
\subfloat{\includegraphics[width=.33\textwidth]{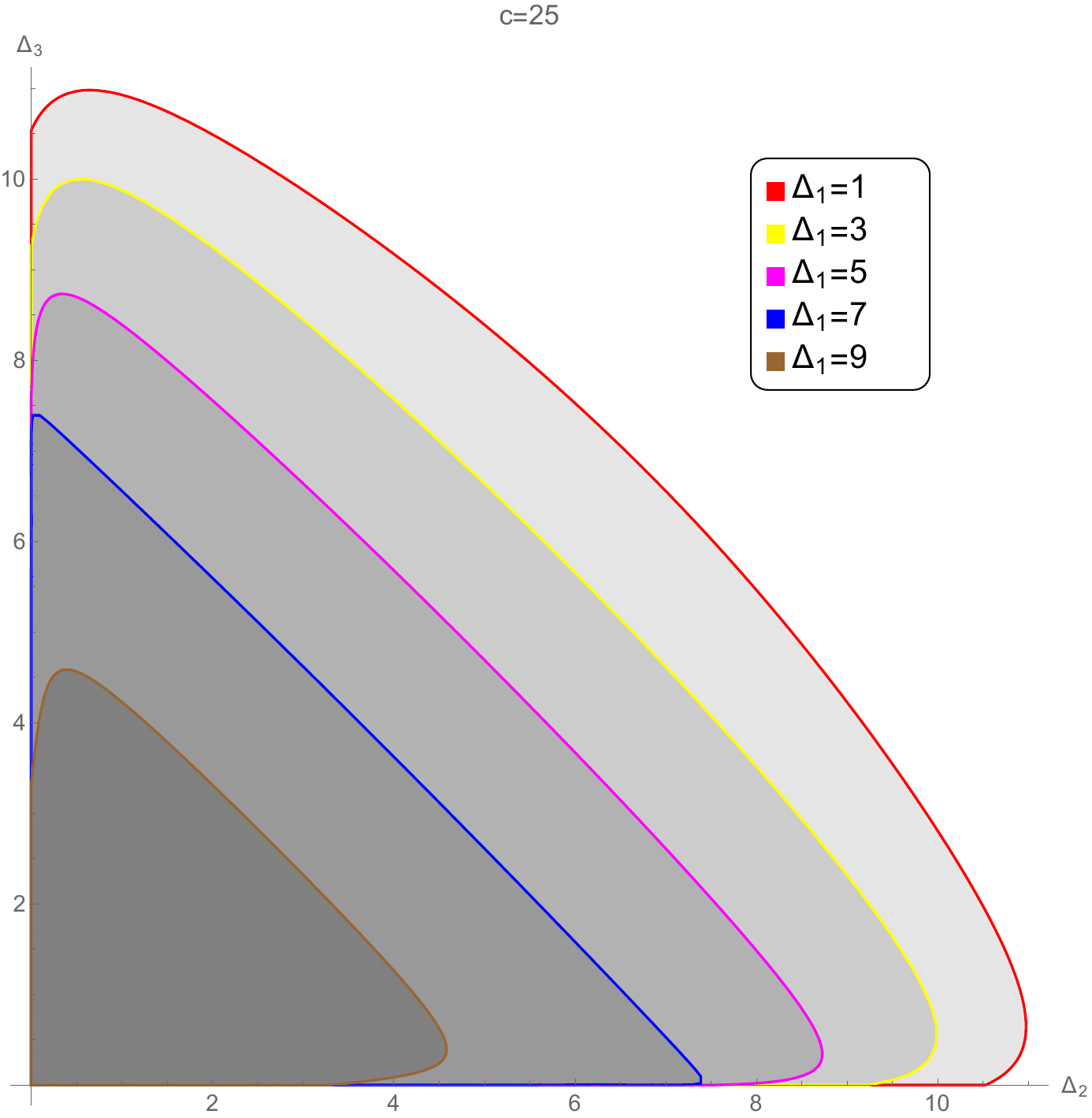}}
\caption{{\bf Top}: Three-dimensional plots of the domain $D^{(3)}_\Delta$ for $c=1,4,25$. {\bf Bottom}: Plots of the cross-sections of these domains for various values of $\Delta_1$.}\label{fig:CriticalDelta}
\end{figure}

For $c>1$, we observe that $W_c(h_1, h_2, h_3)$ is minimized in the limit $h_1=h_2=h_3\to 0$, where it approaches a negative value $-r_c$ (note that in the simultaneous $h_i\to 0$ limit $W_c$ depends on the ratios of the $h_i$'s). For $a_{1,0}$ and $a_{0,1}$ both positive, the domain $D$ bounded by the critical surface $S$ then lies strictly within the domain 
\ie\label{eq:DomainDefinition}
W_c(h_1, h_2, h_3) < {a_{0,1}\over a_{1,0}} r_c,~~~~W_c(\tilde h_1, \tilde h_2, \tilde h_3) < {a_{1,0}\over a_{0,1}} r_c.
\fe
Let us choose $a_{0,1}=a_{1,0}$, and define $\widetilde D$ as the domain $W_c(h_1,h_2,h_3)<r_c$ in $\mathbb{R}^3_{\geq 0}$. Now the compact domain $D_\Delta = \widetilde D+\widetilde D$ (the set of sums of vectors from each set) in $\mathbb{R}^3_{\geq 0}$ may be viewed as a critical domain in the triple of scaling dimensions $(\Delta_1, \Delta_2, \Delta_3)$, with $\Delta_i=h_i+\tilde h_i$, in the sense that structure constants of triples of primaries of dimensions $(\Delta_1,\Delta_2,\Delta_3)$ outside $D_\Delta$ are bounded by those that lie within $D_\Delta$.\footnote{Note that if $\widetilde D$ is convex, then $D_\Delta$ is simply $\widetilde D$ rescaled by a factor of 2, but in fact $\widetilde D$ is generally not convex in the region where one of the weights is small.} Some examples of $D_\Delta$ are shown in Figure \ref{fig:CriticalDelta}.

A subtlety pointed out at the end of section \ref{cbdecompsec} is that the simultaneous $h_i\to 0$ limit of the genus two conformal block with three positive internal weights is distinct from the vacuum block. If we define $W_{c,0}$ to be (\ref{eq:Wc}) computed using the vacuum block, we would find a result that is slightly below $\lim_{h_1=h_2=h_3\to 0^+} W_c(h_1,h_2,h_3)=-r_c$. 
Since we seek critical surfaces such that the structure constants of ``heavy primaries" outside are bounded by those of the ``light primaries" that lie inside the surface, the vacuum block which enters the genus two partition function with coefficient 1 is not relevant, and thus the result (\ref{eq:DomainDefinition}) suffices.

\section{Beyond the $\mathbb{Z}_3$-invariant surface}
\label{beyondz}

In order to write the modular crossing equation for the partition functions on genus two Riemann surfaces of general moduli in a computationally useful manner, we will still work at the $\mathbb{Z}_3$-invariant Renyi surface and expand around the crossing-invariant point $z={1\over 2}$, but with extra insertions of stress-energy tensors $T(z_j)$ and $\tilde T(\bar z_j)$ on any of the three sheets. 

Under the crossing $z\to 1-z$, the transformation of the stress-energy tensors is simple. For instance, it suffices to work with the insertion of $V=L_{-N}\bar L_{-\tilde N}\cdot 1$ on one of the sheets at the point $w$. Here $L_{-N}\equiv L_{-n_1}\cdots L_{-n_k}$ is a Virasoro chain, and $\bar L_{-\tilde N}$ is defined similarly. The crossing transformation sends $V$ to the operator $(-)^{|N|+|\tilde N|}L_{-N}\bar L_{-\tilde N}\cdot 1$ inserted at the position $1-w$. The point $w$ is mapped to the pillow coordinate via (\ref{pillowmap}).
In particular, with $z={1\over 2}$, $\tau=i$, the points $w={1\pm i\over 2}$ are mapped  to $v={\pm 1 + i\over 2}\pi$ (up to monodromies), i.e. the center on the front and back of the pillow.

We can now define the modified conformal blocks with $\widehat{L}_{-R_i}(x)$ insertion on the $i$-th sheet,
\ie\label{deformedBlock}
\mathbb{F}(h_1,h_2,h_3;R_1, R_2, R_3; w|z) &= 3^{-3\sum_{i=1}^3 h_i} \sum_{\{N_i\}, \{M_i\} }  z^{-2h_\sigma + \sum_{i=1}^3(h_i+|N_i|)} w^{\sum_{k=1}^3 (|M_k|-|N_k|-|R_k|)}
\\
&~~
\times \rho( {\cal L}_{-N_3}^\infty h_3, {\cal L}_{-N_2}^1 h_2, {\cal L}_{-N_1}^0 h_1) \rho({\cal L}_{-M_3}^{\infty *} h_3, {\cal L}_{-M_2}^{1*} h_2, {\cal L}_{-M_1}^{0*} h_1)  
\\
&~~
\times \sum_{|P_i|=|N_i|,\; |Q_i|=|M_i| } \prod_{k=1}^3 G^{N_k P_k}_{h_k} G^{M_k Q_k}_{h_k}  \rho( L_{-Q_k} h_k, L_{-R_k}{\rm id}, L_{-P_k} h_k) .
\fe
Here the level sum takes the form of a series expansion in $w$ and $z/w$. For numerical evaluation, it is far more efficient to reorganize the sum as an expansion in $q_1\equiv e^{i(\pi \tau-v)}$ and $q_2\equiv e^{i v}$ instead, where $\tau$ and $v$ are given by (\ref{enome}) and (\ref{pillowmap}).  As is evident from the pillow frame, evaluating at $z={1\over 2}$ and $w={1+i\over 2}$, the effective expansion parameters are $|q_1| = |q_2| = e^{-\pi/2}$, with unit radius of convergence. Explicitly, we have
\ie
{z\over w}= 4 q_2 - 8 q_2^2 + 8 q_1q_2 + 12 q_2^3 - 32  q_1q_2^2 + 4 q_1^2q_2  - 16 q_2^4 +64 q_1q_2^3  - 48 q_1^2q_2^2  + \ldots,
\fe
and $w$ is given by the same series expansion with $q_1$ and $q_2$ exchanged.

For example, the conformal block with a single stress-energy tensor inserted in the first sheet, up to total level 2 in $q_1$ and $q_2$, is given by
\ie{}
&\mathbb{F}(h_1,h_2,h_3;R_1=\{2\},R_2= \varnothing,R_3= \varnothing; q_1,q_2)~~~~~~~~~~~~~~~~~~~~~~~~~~~~~~~~~~~~~~~~~~~~~~~~~~~~~~~~~~~~~~~~~~~~~\\
&=\left({16 q_1q_2\over27}\right)^{h_1+h_2+h_3}{(16q_1q_2)^{-{2\over9}c}\over q_1^2}\bigg\{{h_1\over 16}+{1\over 36}\bigg(18h_1 q_1 -i\sqrt{3}(h_2-h_3)(q_1-q_2)\bigg)\\
&~~~+{1\over 216 h_1 h_2 h_3}\bigg[h_1 h_2 h_3 q_1^2 \left(8 c+413 h_1 -4(h_2+h_3)-48i\sqrt{3}(h_2-h_3)\right)\\
&~~~+4 q_2 q_1 \bigg(h_1^2 \left(6 (c+9) h_3 h_2+h_2^3+h_3^3\right)+ h_1^4\left(h_2+h_3\right)-2h_1^3 \left(h_2^2+h_3^2\right) \\
&~~~+h_1h_2 h_3 \left(h_2-h_3\right) \left(h_2-h_3+12 i \sqrt{3}\right) +h_2 h_3 \left(h_2-h_3\right)^2\bigg)+h_1 h_2 h_3 q_2^2 \left(8 c+35 h_1-4 h_2-4 h_3\right)\bigg]\\
&~~~+\ldots\bigg\}.
\fe
If we symmetrize (\ref{deformedBlock}) with respect to $R_1, R_2, R_3$, we recover the conformal block of the $\mathbb{Z}_3$-invariant Renyi surface considered in the previous section, differentiated with respect to $z$, up to a conformal anomaly factor. In particular, summing over insertions of a single stress-energy tensor on one of the three sheets, we find
\ie
&\mathbb{F}(h_1, h_2, h_3; R_1=\{2\},R_2= \varnothing,R_3= \varnothing; q_1, q_2)+
(2~{\rm cyclic~permutations~on~}R_1, R_2, R_3)
\\
&=C(q_1,q_2)\partial_q\mathcal{F}(h_1, h_2, h_3; q)\big |_{q=q_1q_2}+cB(q_1,q_2)\mathcal{F}(h_1, h_2, h_3; q=q_1q_2),
\fe
where the first term on the RHS is due to deformation of the modulus $z$ or $q$ and the second term is due to the conformal anomaly (from a Weyl transformation that flattens out the pillow geometry after the insertion of the stress-energy tensor). The functions $C$ and $B$ are independent of $h_i$ and $c$; they admit series expansions in $q_1$ and $q_2$ of the form
\ie{}
C(q_1,q_2)=&{q_2\over16q_1}+{q_2\over2}+{q_2\over8}\left(15q_1+8q_2+{q_2^2\over q_1}\right)+{1\over 2}q_2\left(9q_1^2 + 16 q_1 q_2 + 3 q_2^2\right)+\ldots,
\\
B(q_1,q_2)=&{1\over72q_1^2}+{1\over9q_1}+{19q_1^2+4q_1q_2+5q_2^2\over36q_1^2}+{17q_1^2 + 8 q_1 q_2 + 11 q_2^2\over 9q_1}+\ldots.
\fe

A complete set of genus two modular crossing equations can now be written as
\ie\label{crossgen}
&  (-)^{\sum_{j=1}^3 (|R_j|+|\tilde R_j|)} \sum_{(h_i, \tilde h_i)} C_{h_1, h_2, h_3; \tilde h_1, \tilde h_2, \tilde h_3}^2 \mathbb{F}(h_1,h_2,h_3;R_1, R_2, R_3;w|z) \mathbb{F}(\tilde h_1,\tilde h_2,\tilde h_3;\tilde R_1, \tilde R_2, \tilde R_3;\bar w |\bar z)
\\
&= \sum_{(h_i, \tilde h_i)} C_{h_1, h_2, h_3; \tilde h_1, \tilde h_2, \tilde h_3}^2 \mathbb{F}(h_1,h_2,h_3;R_1, R_2, R_3;1-w|1-z) \mathbb{F}(\tilde h_1,\tilde h_2,\tilde h_3;\tilde R_1, \tilde R_2, \tilde R_3; 1-\bar w|1-\bar z) .
\fe
If we take into account all possible choices of integer partitions $R_j$ and $\tilde R_j$, it suffices to evaluate this equation at the crossing-invariant point $z=\bar z={1\over 2}$, with the choice $w={1+i\over 2}$, $\bar w={1-i\over 2}$. The consequence of (\ref{crossgen}) in constraining structure constants in unitary CFTs is currently under investigation.

\section{Discussion}
\label{discussion}

The main results of this paper are the formulation of genus two modular crossing equations in an explicitly computable manner, by working on the Renyi surface as well as expanding around it. As an application, we found compact critical surfaces that bound domains $D \subset \mathbb{R}^3_{\geq 0}$ such that structure constants $C_{ijk}$ involving a triple of primaries whose dimensions $(\Delta_i, \Delta_j,\Delta_k)$ or twists $(\tau_i, \tau_j, \tau_k)$ are outside of $D$ are bounded by those that lie within $D$. The existence of the compact critical surface is a nontrivial consequence of genus two modular invariance that does not follow easily from the analysis of individual OPEs: roughly speaking, the crossing equation for the sphere 4-point function bounds light-light-heavy structure constants in terms of light-light-light ones, but the genus two modular crossing equation also bounds light-heavy-heavy and heavy-heavy-heavy structure constants in terms of light-light-light ones.

In deriving the critical surface, we have used merely a tiny part of the genus two crossing equation, namely the first order $z$ and $\bar z$ derivatives of the Renyi surface crossing equation evaluated at the crossing invariant point $z=\bar z = {1\over 2}$. Clearly, stronger results for the critical surfaces (that bound smaller domains) should be obtained by taking into account higher order $z$ and $\bar z$ derivatives of the crossing equation. This is rather tricky to implement numerically through semidefinite programming, simply due to the fact the genus two conformal block decomposition involves 3 continuously varying scaling dimensions and 3 spins. To implement the crossing equation through \cite{Simmons-Duffin:2015qma}, for instance, one may attempt to vary the sum of the 3 scaling dimensions, and sample over their differences as well as truncating on the spins, but such a sampling would involve a huge set of conformal blocks that is hard to handle numerically. At the moment this appears to be the main technical obstacle in optimizing the genus two modular bootstrap bounds.\footnote{A potentially more efficient numerical approach would be based on sum-of-squares optimization, as is explained to us by D. Simmons-Duffin.}

Many more constraints on the structure constants $C_{ijk}$ can in principle be obtained by consideration of higher order derivatives of the genus two crossing equation. For instance, combining first and third order derivatives, analogously to \cite{Rattazzi:2008pe, Hellerman:2009bu}, one can deduce the existence of structure constants $C_{ijk}$ with say the dimensions $(\Delta_i, \Delta_j, \Delta_k)$ lying within a small domain (typically, such a domain is strictly larger than one that is bounded by a critical surface). The genus two modular invariance potentially has the power to constrain CFTs with approximately conserved currents (i.e. primaries with very small twist): if such a current operator propagates through one of the three handles of the genus two surface, modular invariance should constrain the pairs of operators propagating through the other two handles according to representations of an approximate current algebra or $W$-algebra. Typically, when OPE bounds or (genus one) modular spectral bounds are close to being saturated \cite{Collier:2016cls}, one finds that there are necessarily low twist operators in the spectrum. For instance, this strategy may be used to severely constrain (and possibly rule out) unitary compact CFTs with central charge $c$ slightly bigger than 1.

There is another genus two conformal block channel (the ``dumbbell channel") that we have not discussed so far, namely one in which the genus two surface is built by plumbing together a pair of 1-holed tori. The conformal block decomposition of the genus two partition function in this channel involves the torus 1-point functions, or the structure constants $C_{ijj}$ where a pair of primaries are identified. The modular covariance of the torus 1-point function cannot be used by itself to constrain $C_{ijj}$ in a unitary CFT, since $C_{ijj}$ does not have any positivity property in general. In the dumbbell channel decomposition of the genus two partition function, the structure constants appear in the combination $C_{ijj} C_{ikk}$, allowing for the implementation of semidefinite programming. In our present approach via expansion around the Renyi surface, it appears rather difficult to perform the conformal block decomposition in the dumbbell channel explicitly. How to incorporate this channel in the genus two modular bootstrap is a question left for future work.

\section*{Acknowledgements}

We are grateful to Alex Maloney for pointing out to us the significance of the $\mathbb{Z}_3$-invariant Renyi surface and for sharing his notes on the subject. We would also like to thank Ying-Hsuan Lin and David Simmons-Duffin for discussions and comments on a preliminary draft. XY thanks Simons Collaboration Workshop on Numerical Bootstrap at Princeton University and ``Quantum Gravity and the Bootstrap" conference at Johns Hopkins University for their hospitality during the course of this work. This work is supported by a Simons Investigator Award from the Simons Foundation and by DOE grant DE-FG02-91ER40654. MC is supported by Samsung Scholarship. SC is supported in part by the Natural Sciences and Engineering Research Council of Canada via a PGS D fellowship. Explicit computations of the conformal blocks in this work were performed with Mathematica using the Weaver package \cite{Whalen:2014fta}.

\bibliographystyle{JHEP}
\bibliography{g2}

\end{document}